\documentclass[twocolumn, aps, groupedaddress, superscriptaddress, nofootinbib]{revtex4-1}
\usepackage[utf8]{inputenc}
\usepackage{braket}
\usepackage{amsmath}
\DeclareMathOperator{\Tr}{Tr}
\usepackage{dcolumn}   
\usepackage{bm}        
\usepackage{amssymb}
\usepackage{mathtools}
\usepackage{tikz}
\usepackage{qcircuit}
\usepackage{appendix}
\usepackage{hyperref}
\usepackage{subcaption}
\usepackage{amsmath,amsfonts,amssymb}
\usepackage{mathtools}

\usetikzlibrary{arrows.meta}

\def\BibTeX{{\rm B\kern-.05em{\sc i\kern-.025em b}\kern-.08em
    T\kern-.1667em\lower.7ex\hbox{E}\kern-.125emX}}

\begin{document}
\title{A Unified Framework for Quantum Supervised Learning}
\author{Nhat A. Nghiem} 

\affiliation{Department of Physics and Astronomy, State University of New York at Stony Brook, Stony Brook, NY 11794-3800, USA}
\author{Samuel Yen-Chi Chen}
\affiliation{Computational Science Initiative, Brookhaven National Laboratory, Upton, NY 11973, USA}

\author{Tzu-Chieh Wei}
\affiliation{C. N. Yang Institute for Theoretical Physics, State University of New York at Stony Brook, Stony Brook, NY 11794-3840, USA}
\affiliation{Department of Physics and Astronomy, State University of New York at Stony Brook, Stony Brook, NY 11794-3800, USA}
\begin{abstract}
 Quantum machine learning is an emerging field that combines machine learning with advances in quantum technologies. Many works have suggested great possibilities of using near-term quantum hardware in supervised learning. Motivated by these developments, we present an embedding-based framework for supervised learning with trainable quantum circuits. We introduce both explicit and implicit approaches. The aim of these approaches is to map data from different classes to separated locations in the Hilbert space via the quantum feature map. We will show that the implicit approach is a generalization of a recently introduced strategy, so-called \textit{quantum metric learning}. In particular, with the implicit approach, the number of separated classes (or their labels) in supervised learning problems can be arbitrarily high with respect to the number of given qubits, which surpasses the capacity of some current quantum machine learning models. Compared to the explicit method, this implicit approach exhibits certain advantages over small training sizes. Furthermore, we establish an intrinsic connection between the explicit approach and other quantum supervised learning models. Combined with the implicit approach, this connection provides a unified framework for quantum supervised learning. The utility of our framework is demonstrated by performing both noise-free and noisy numerical simulations. Moreover, we have conducted classification testing with both implicit and explicit approaches using several IBM Q devices.
\end{abstract}
\maketitle

\section{Introduction}
Quantum computation has been intensively studied over the past few decades and is expected to outperform its classical counterpart in certain computational tasks~\cite{nielsen2002quantum, harrow2017quantum,arute_quantum_2019}. In this novel approach for computation, information is stored in the quantum states of an appropriately chosen and designed physical system, which resides in a complex Hilbert space $\mathcal{H}$, and quantum bits (qubits) are used as the underlying building blocks and processing units. The power of a quantum computer is in its ability to store and process information coherently in the tensor-product Hilbert space~\cite{nielsen2002quantum} with entanglement being a characteristic byproduct or even a potential resource for quantum information processing~\cite{briegel2009measurement,raussendorf2012quantum}. Quantum computations have been shown to provide dramatic speedup in solving some important computational problems, such as factorization of a large number via Shor's algorithm~\cite{shor1999polynomial} and the unstructured search using Grover's algorithm~\cite{grover1996fast}, which are two prominent examples among many that have been discovered. 

At the same time, machine learning (ML) has become a powerful tool in modern computation. For example, ML has been successful in computer vision \cite{Simonyan2014VeryRecognition, Szegedy2014GoingConvolutions, Voulodimos2018DeepReview}, natural language processing \cite{Sutskever2014SequenceNetworks}, and drug discovery \cite{vamathevan2019applications}. Building on this history, a natural application of quantum computers also may provide substantial speedup~\cite{biamonte2017quantum, wittek2014quantum, schuld2019machine, dunjko2018machine}.  
Several previous works have revealed potential quantum advantages in the field of unsupervised learning~\cite{aimeur2013quantum, otterbach2017unsupervised, wiebe2014quantum, lloyd2013quantum, kerenidis2019q}. For example, in Ref.~\cite{lloyd2013quantum}, the authors provide quantum algorithms for clustering problems, which could, in principle, yield an exponential speedup. In Ref.~\cite{kerenidis2019q}, the authors introduce a quantum version of $k$-means clustering, namely, $q$-means, and present an efficient quantum procedure. 

Using quantum computation in supervised learning also has garnered increased attention~\cite{schuld2018supervised, benedetti2019parameterized, sergioli2019new}. For example, the authors of Ref.~\cite{rebentrost2014quantum} present a quantum version of support vector machines (SVM) that showed possible exponential speedup. For near-term applications, variational strategies have been proposed to classify real-world data~\cite{farhi2018classification, schuld2018supervised, schuld2020circuit, havlivcek2019supervised, mitarai2018quantum}. Classification is among the standard problems in supervised learning \cite{lecun2015deep, krizhevsky2012imagenet}, and variational methods using short-depth quantum circuits with trainable parameters
have given rise to a quantum-classical hybrid optimization procedure. Such frameworks have proven to be capable of performing complex classification tasks \cite{farhi2018classification,mitarai2018quantum,schuld2020circuit,grant2018hierarchical, liu2019hybrid, lu2020quantum}, and many more likely will appear. It is probable that such variational methods will be able to learn complex representations while still being robust to noise in near-term quantum devices (e.g., noisy intermediate-scale quantum [NISQ]) \cite{mitarai2018quantum, du2020learnability, huang2020experimental}. 

``Traditional'' quantum supervised learning (QSL) models rely on the encoding of classical data $x$ into some quantum state $\ket{\psi(x)}$. This state then undergoes a parameterized quantum circuit $U(\theta)$. At the end, the state is measured. The outcome of the measurement usually is interpreted as the output of the learning model. Although the procedures of previously proposed works~\cite{mitarai2018quantum, havlivcek2019supervised, schuld2020circuit, schuld2019machine} appear similar, the motivation underlying their strategies seems varied. For example, the quantum circuit learning algorithm proposed in~\cite{mitarai2018quantum} is inspired by classical neural networks. Meanwhile, in \cite{schuld2019quantum, havlivcek2019supervised}, the authors exploit and formally establish the connection between quantum computation and the kernel method, where they interpret the step of encoding classical data $x$ into the quantum state as a quantum feature map. Thus, a clear picture emerges: classical data $x$ are embedded into some quantum state $\ket{x}$, i.e., a data point in Hilbert space $\mathcal{H}$. (This space also is called the \textit{quantum feature space}, analogous to the feature space in classical ML.) Then, a decision boundary is learned by training the variational circuit to adapt the measurement basis, which is analogous to the classical approach where a decision boundary is learned to separate classes.

Metric learning is a well-known method in the classical ML context~\cite{chopra2005learning}. The aim is to learn an appropriate distance function over data points. This method recently has been extended to the quantum context by Lloyd et al. (\cite{lloyd2020quantum}). Instead of focusing on training the variational layer that adapts the measurement basis, the authors propose to train the embedding circuit and proffer a remarkable notion of ``well-separation'' of data points. Per their argument, a significant amount of computational power spent on processing classically embedded data can be eased using such a strategy.

Aside from such an advantage, we pose that the ability to use a quantum circuit to represent data in a complex Hilbert space and the idea of ``well-separation'' have further remarkable consequences. We argue that previous ``traditional'' QSL methods \cite{mitarai2018quantum, schuld2020circuit, schuld2019machine, havlivcek2019supervised} essentially achieve certain ``well-separation'' of data points. Here, we provide a unified, generic framework and categorize approaches to two different types, \textit{implicit} and \textit{explicit}, which are described in detail, backed up by numerical simulations, and tested on real quantum devices. The goal is to train the embedding circuit to produce clusters of data from different classes. In the {\it implicit} approach, the ``centers'' of these clusters are random. In the {\it explicit} approach, the cluster ``centers'' are constrained to lie in or nearby some predetermined subspaces of the Hilbert space. We show that both explicit and implicit approaches exhibit promising classification ability. Particularly, the method proposed by Lloyd et al.~(\cite{lloyd2020quantum}) is a binary version of the implicit approach. We point out that the explicit approach can conceptually unify ``traditional'' QSL methods, such as~\cite{mitarai2018quantum, schuld2019machine, schuld2020circuit}. These two approaches then constitute our unified framework for QSL.

%
\medskip 
The following summarizes the contributions of this work: 
\begin{itemize}
    \item We introduce two approaches for QSL, implicit and explicit, that constitute a generic embedding-based  framework. 
    
    \item We show that the implicit approach is the generalization of the metric quantum learning method proposed in \cite{lloyd2020quantum}. Such generalization allows us to manage the multi-class classification problem. The number of separated classes (or labels) is independent of qubits used in the quantum circuit. Therefore, it sheds light on constructing a universal quantum classifier.
    
    \item We demonstrate that the explicit approach can conceptually unify other models for QSL. Along with the generalization provided by the implicit approach, our work provides a complete unification of QSL frameworks.
    
    \item We implement both learning approaches on NISQ devices and compare the results with noisy simulations. We demonstrate the framework's success and clarify the cases where the results on real devices and noisy simulations do not agree well. 
    
   
\end{itemize}
The structure of the paper is as follows: Section \ref{sec:sec2} presents the main conceptual tool of our framework. In Sections \ref{sec: implicit} and \ref{sec: explicit}, we discuss the implicit and explicit approaches, present results from numerical experiments and runs from real devices, and provide numerical evidence that the implicit approach is especially robust with small training size. Some discussions regarding our framework's prospects in the near-term era are presented in Section~\ref{sec: prospect}. Section~\ref{conclude} concludes the primary work. Appendix~\ref{sec: unification} provides an additional example to illustrate the unification. Appendix~\ref{caveatovall} discusses a remarkable consequence of focusing on the embeddings part instead of the measuring part in QSL, which could avoid a systematic issue of misclassification of the one-versus-all strategy.

\section{General framework}

\subsection{Basic concept}
\label{sec:sec2}
We first introduce the basic concept of classification, which can be illustrated by a simple map: 
\begin{equation}
\label{eqn: metricmap}
x \longrightarrow \vec{f}(x, \theta), 
\end{equation}
where 
\begin{equation}
\vec{f}(x, \theta) = \begin{bmatrix}
f_0 \\
\vdots \\
f_i \\
\vdots \\
f_{L-1} \\
\end{bmatrix}. \\
\end{equation}
$\in \mathbb{R}^L$ is called a classifying vector of some input data $x$, and $\theta$ refers to the network's or circuit's parameters. $\{f_i\}$ is generally of the form: \\
\begin{equation}
    f_i =  \langle x\rvert\mathcal{M}_i\lvert x\rangle
= \Tr{ (\ket{x}\bra{x} \mathcal{M}_i ), }  
\end{equation}
where $\ket{x}$ is the corresponding quantum state of classical data x, and $\mathcal{M}_i $, in general, is some Hermitian operator. We generally assume that N-dimensional classical data x is mapped to $\ket{x}$ via a k-qubits parameterized circuit. The specific formula for $\mathcal{M}_i$ depends on either the implicit or explicit approach (to be discussed later).
The value of $f_i$ depends on circuit parameters $\theta$ and the input feature $x$. \\

In the supervised learning problem with $L$ separated labels, we are given a training set together with corresponding labels $X \times Y = \{x, i\} $, where $i \in \{0,1,..,L-1\}$ is the label of the data point $x$. 
\begin{figure}[htbp]
    \includegraphics[scale = 0.3]{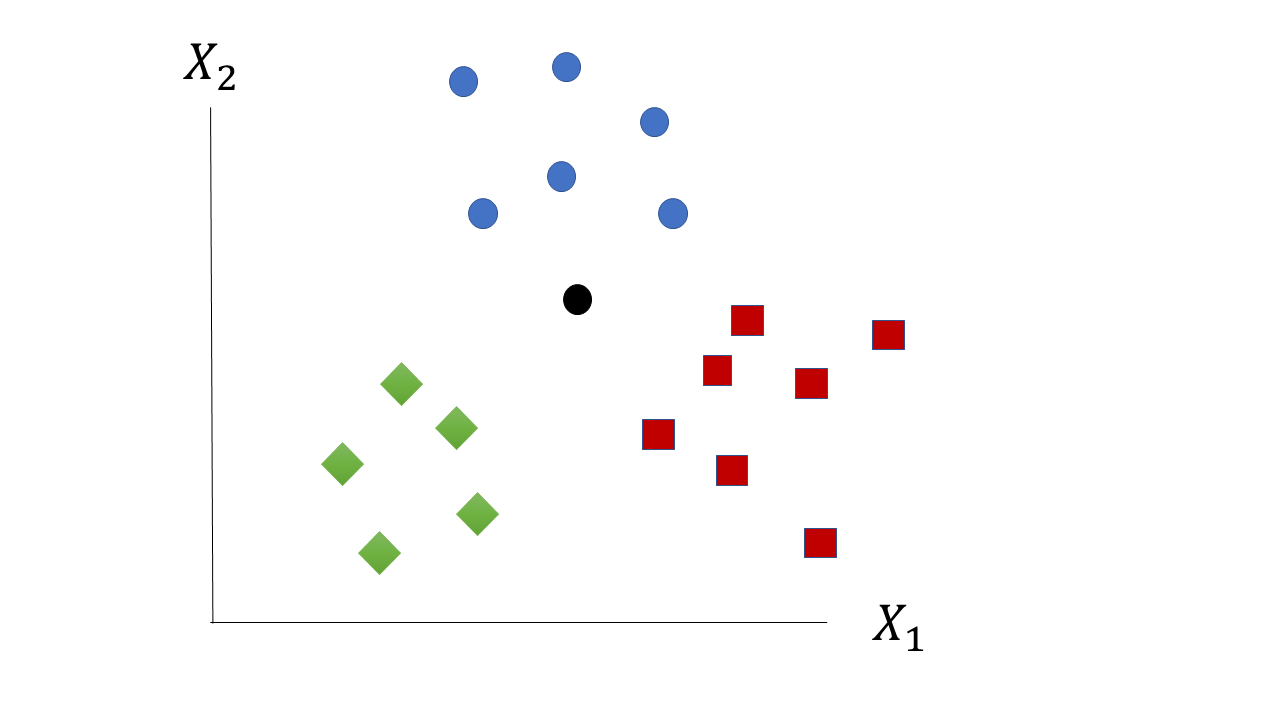}
    \caption{An example supervised learning problem with $L = 3$ labels. The black circle is some unseen data. }
    \label{fig: svl}
\end{figure}
We need to predict the label for some other unseen data $x$. In the quantum setting, we simply use its representation by a quantum state $\ket{x}$ instead of the classical data $x$. The number of components $\{f_i\}$, or equivalently the dimension of $\vec{f}(x,\theta)$, is denoted by $L$. The value $f_i$ (for convenience, assumed to be in the range $0 \leq f_i \leq 1$) quantifies the likelihood that some input $x$ have any of labels $\{i\}$. In this sense, the method is somehow similar to the classical neural network, where the information is fed forward from the input layer to the output layer. There have been numerous works that  explore the relation between quantum computation and neural networks \cite{schuld2020circuit,farhi2018classification,schuld2014quest,rebentrost2018quantum, cong2019quantum, schuld2014quest, beer2020training}. The key relation extends from the building block of the quantum circuit model: quantum gates. These gates carry out unitary transformation on the input quantum state, which is a vector in some Hilbert space ${\cal H}$. In the graphical representation (distinctively illustrated in Ref.~\cite{schuld2020circuit}), the action of a quantum gate on the input state $\ket{\psi}$ produces an output state $\ket{\rho}$ and can be represented as a fully connected two-layer network. Hence, a full quantum circuit generally can be represented by such a fully connected network with a certain number of layers. Measuring quantum states then corresponds to a non-linear activation function.

Thus, to make prediction, we ``forward'' $x$ to such a classifying vector $\vec{f}(x,\theta)$ and assign to it a label according to the highest value of $\{f_i\}$. The accuracy of correct assignment depends on circuit parameters $\theta$. Now, we provide a strategy to train the circuit. For each label $i$, assume there are $N_i$ training points with such a label, and there are a total of $N$ data points, where $N=\sum_i N_i$. Let $\vec{y_i}$ be the real, so-called \textit{label vector}, of class $i$ (which has dimension $L$) with components $\{ y_i^j \}_{j=1}^L = \delta_{ij}$, where $\delta_{ij}$ is the Kronecker delta function. Let $\vec{f}_i^j $ be the classifying vector of the $j$-th data. We minimize the following cost or loss function: \begin{equation}
\label{eqn: cost}
    C = \frac{1}{L}\sum_{i=1}^L
    \frac{1}{N_i}\sum_{j=1}^{N_i}  |\vec{f}(x_i^j,\theta)- \vec{y_i} |,
\end{equation}
where $x_i^j$ is the $j$-th data point in class $i$. We finally note that any reasonable form of the loss function should work. \\

\textit{State Overlaps:} Given two {\it pure} quantum states represented by density matrices $\rho\equiv|\rho\rangle\langle\rho|$ and $\phi \equiv |\phi\rangle\langle\phi|$, the overlaps, i.e., a similarity measure, on these two states is given by $\Tr(\rho\phi)=|\langle\rho|\phi\rangle|^2$. State overlaps play an important role in our subsequent construction of the implicit and explicit approaches.
In the general case of mixed states, the SWAP test quantum procedure \cite{nielsen2002quantum} can be used to evaluate $\Tr(\rho\phi)$ up to an additive error $\epsilon$. In the special case of pure states, the \textit{inversion test} \cite{lloyd2020quantum} can be used to evaluate the overlaps between two quantum states $|\bra{\phi}\cdot\ket{\rho}|^2 $, provided the circuit $U$ to create either $\ket{\phi}$ or $\ket{\rho}$ can be efficiently inverted. Both schemes require only shallow circuits. In our subsequent experiments, we also will implement both schemes for classification on real devices.

\subsection{Implicit Approach}
\subsubsection{Construction}
\label{sec: implicit}
\begin{figure}[t]
    \includegraphics[width = \linewidth]{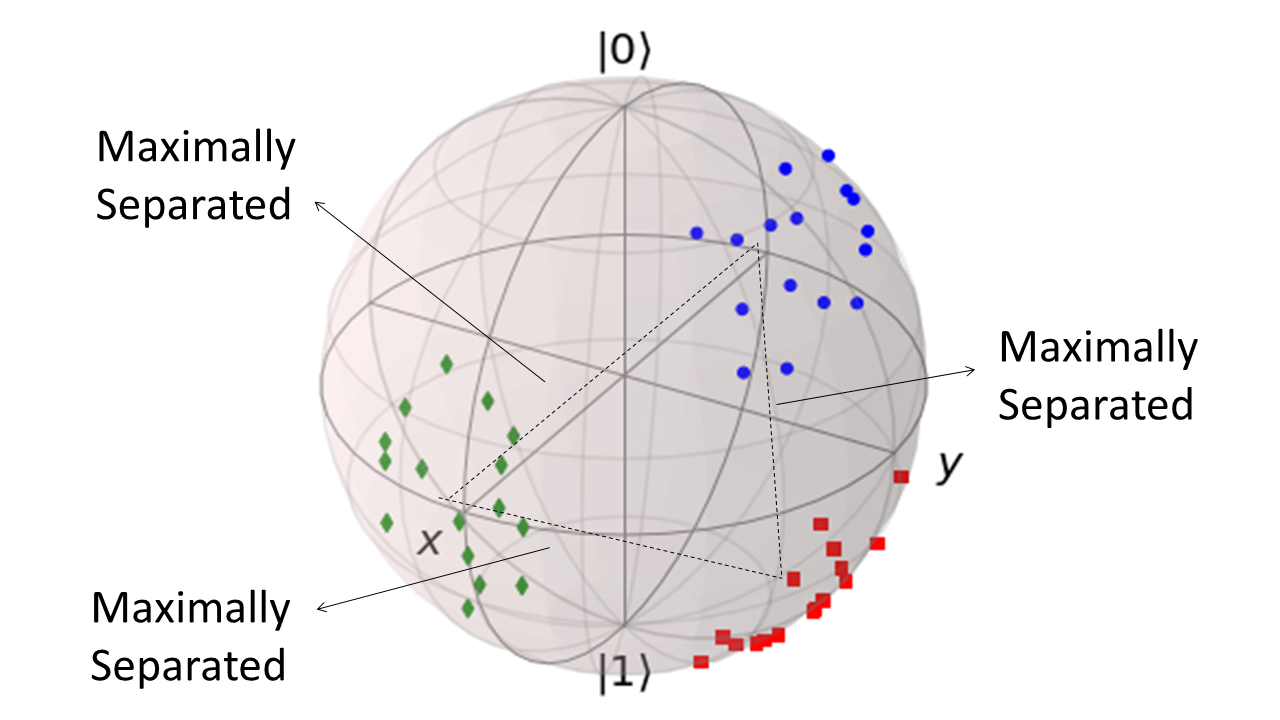}
    \caption{ \textbf{Illustration of the implicit approach}. After the training procedure, the ``distance'' between any clusters (represented by dotted lines) becomes maximal. The ``center'' of each cluster is not fixed as the training process will move them to produce maximally separated clusters. \textit{Note}: the data points in this picture are not related to the Iris dataset in our subsequent experiment or the data points in Fig.~\ref{fig: explicitbloch}  }
    \label{fig: implicitbloch}
\end{figure}

In this approach, the data from the same class, after going through the quantum circuit $\Phi(x, \theta)$, produce clusters (closed data points) in the Hilbert space $\mathcal{H}$. Clusters corresponding to different classes should become maximally separated after minimizing the cost function (see Fig.~\ref{fig: svl} \& \ref{fig: implicitbloch}). 

To describe our supervised learning problem, we assume that for each label $i$, there are $N_i$ training points that will be transformed to quantum states $\{ \ket{x_i^j}\}_{j=1}^{N_i}$. 
The formula for $\mathcal{M}_i$ in this case is: \\
\begin{equation}
    \mathcal{M}_i = \frac{1}{N_i} \sum_j \ket{x_i^j}\bra{x_i^j,} 
\end{equation}
which is exactly an ensemble of quantum states: $\sigma_i =\frac{1}{N_i} \sum_j \ket{x_i^j}\bra{x_i^j} $. This ensemble may be interpreted as the collection of the corresponding training points from class $i$ on $\mathcal{H}$, and it can be obtained by sampling from the training set $\{x_i^j\}\vert_{j=1}^{N_i}$. The classifying vector $\vec{f}(x,\theta)$ now becomes: 
\begin{equation}
\label{eqn: metricimplicit  }
\begin{bmatrix}
\Tr{(\ket{x}\bra{x} \sigma_0  ) } \\
\Tr{(\ket{x}\bra{x} \sigma_1  ) } \\
.\\
.\\
\Tr{(\ket{x}\bra{x} \sigma_{L-1}  ) } \\ 
\end{bmatrix}.
\end{equation}
We will focus on optimizing those quantities $\{f_i\}$ and using the values to assign a label to $x$. 

After applying Eq.~(\ref{eqn: cost}), the cost function  becomes
\begin{equation}
    \label{eqn: newcost}
    C =  1 - \frac{1}{L} \sum_{i=1}^L \Tr{\sigma_i^2} + \frac{2}{L} \sum_{i< j} \Tr{\sigma_i\sigma_j}. 
\end{equation}
We note that each cross terms
\begin{equation}
\label{eqn: pairoverlap}
    \Tr(\sigma_i\sigma_j) =  \frac{1}{N_i N_j} \sum_{k=1}^{N_i}\sum_{p=1}^{N_j} | \bra{x_i^k}\cdot\ket{x_j^p} |^2
\end{equation}
is a sum of the modulus square of overlaps.
Hence, in the training procedure, we can use either the SWAP or inversion test to evaluate the cost.  

Consider a binary classification problem ($L=2$). The cost function is: 
\begin{equation}
    C = 1 - \frac{1}{2}(\Tr{\sigma_1^2} +\Tr{\sigma_2^2}) + \Tr{\sigma_1\sigma_2} = 1- \frac{1}{2} \Tr{ (\sigma_1-\sigma_2)}^2.  
\end{equation}

\begin{figure}[t]
    \includegraphics[width = \linewidth]{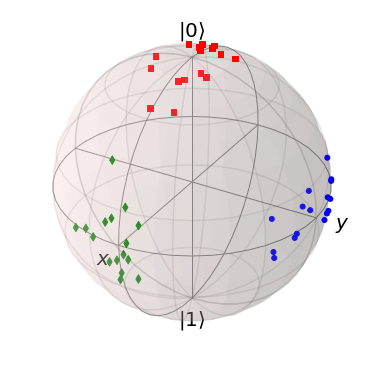}
    \caption{ \textbf{Illustration of the explicit approach}. The ``center'' of each cluster is fixed. For class 0 (red points), the position of the center is $v_0$ = (0,0,1). For class 1 (green points), the position is $v_1$ = (1,0,0). For class 2, the position is $v_2$ = (0,1,0). Notably, this separation is not exactly the same as desired in Eq.~(\ref{eqn: decomposition}) because the Hilbert space associated with a single qubit has dim = 2, and it only can be decomposed into, at most, two orthogonal subspaces. Nevertheless, this figure illustrates the idea of the explicit approach.  }
    \label{fig: explicitbloch}
\end{figure}
The optimization will be minimizing the Hilbert-Schmidt distance between two data clusters, which is highlighted in Ref.~\cite{lloyd2020quantum}. Therefore, we have shown this implicit approach is a generalization of the binary method discussed in Ref.~\cite{lloyd2020quantum}.

\smallskip
\subsubsection{Training with QRAM}
If the quantum random access memory (QRAM)~\cite{giovannetti2008quantum} is available, the cost for the training and testing procedure will be reduced by a factor of $\sim$ $\sum_{i\leq j}N_iN_j$ and $\sum_i N_i$, respectively, where $N_i$ is the number of the data points in each training set $i$. The calculation of cost function and data classification can be done in time $\mathcal{O}(1)$. The data can be loaded corresponding to class $i$ to a quantum state $\ket{\psi_i} = \frac{1}{\sqrt{N_i}}\sum_{j=1}^{N_i} \ket{ ``{x}_i^j"}\ket{j}$, where the index $j$ represents the address of the memory where ${x}_i^j$ is residing. We use $\ket{ ``{x}_i^j"}$ to denote the classical data (not the embedded feature state) loaded to a quantum register. 

A third register is initialized in $|0\dots0\rangle$ and is used to implement the quantum feature. The application of $R_y(x)$ (refer to Fig.~\ref{fig: QAOA}) on this register with its argument being $x_i^J$ can be done by performing the rotation conditioned on the first register with classical values ${x}_i^j$, i.e., a conditional rotation $c-R_y(x)$. In this way, at the expense of a more complicated circuit to implement the conditional rotation, we obtain the entangled state
$\ket{\psi'_i} = \frac{1}{\sqrt{N_i}}\sum_{j=1}^{N_i}\otimes \ket{ ``{x}_i^j "}\otimes\ket{j}\otimes |\Phi(x_i^j,\theta)\rangle$. Tracing over the first and second registers (i.e., without doing anything on them afterwards), the third register is in the state $\sigma_i=\sum_j|\Phi(x_i^j,\theta)\rangle\langle \Phi(x_i^j,\theta)|/N_i$.
With the QRAM, we do not need to repeat the circuits (with different rotations) to sample every data point individually from the training set to obtain an effective $\sigma_i$ (about $N_i$ times).

Using the controlled-SWAP gate on this third register and another other embedded state $|x\rangle$ for an unknown data point (i.e., the SWAP test), we can directly measure their fidelity $\langle x|\sigma_i|x\rangle$.  However, computing the pairwise overlaps in Eq.~(\ref{eqn: pairoverlap}) without the QRAM will require using the SWAP test $N_iN_j$ times.
As such, it would be useful to design an efficient quantum subroutine of low-depth circuits to evaluate the cost in Eq.~(\ref{eqn: newcost}), directly exploiting the QRAM and reducing the iterative evaluation steps.

\subsubsection{Classifying over a large number of classes }
In most current quantum ML models, the measurement outcome usually is interpreted as the outcome of learning models. In a binary classification problem, one-qubit measurement suffices to classify the data as there are only two possible outcomes, and one can draw inferences from such a measurement. For example, if the probability of obtaining class zero $\mathcal{P}$(outcome = 0) $\geq$ 0.5, we then assign the data to class 0. Otherwise, we assign it to class 1. For multi-class classification, multi-qubit measurement needs to be employed. A circuit with $k$ qubits can classify up to $2^k$ different labels. Our implicit approach surpasses this because we only aim to get the classifying vector $\vec{f}$. The dimension of $\vec{f}$, or the number of classes, can be quite large. Real-world supervised learning problems may contain overwhelmingly numerous classes, such as in face recognition. Thus, our framework may prove useful for these practical tasks.

Still, even with a single qubit, multi-classification can be done using this approach (see Fig.~\ref{fig: implicitbloch}). Relevant work has been carried out in Ref.~\cite{perez2020data}, showing that a single qubit is sufficient to construct a universal classifier and is able to deal with multidimensional input data and multi-label output of supervised learning problems. To handle a multidimensional input, the authors propose a data re-uploading strategy. They achieve the multi-classification by introducing bias term $\lambda$ in the measurement outcome. Our approach differs from~\cite{perez2020data} because we use the parameterized quantum circuit to represent the data in the Hilbert space and exploit its ``vastness.'' Data from different classes are ``aligned'' in separate locations. To handle multidimensional data, one may opt to follow the same strategy as in Ref.~\cite{perez2020data} or engage a different embedding routine. Addressing the problem of efficient data encoding is beyond the scope of this work.

\begin{figure}[t]
    \includegraphics[scale = 0.28]{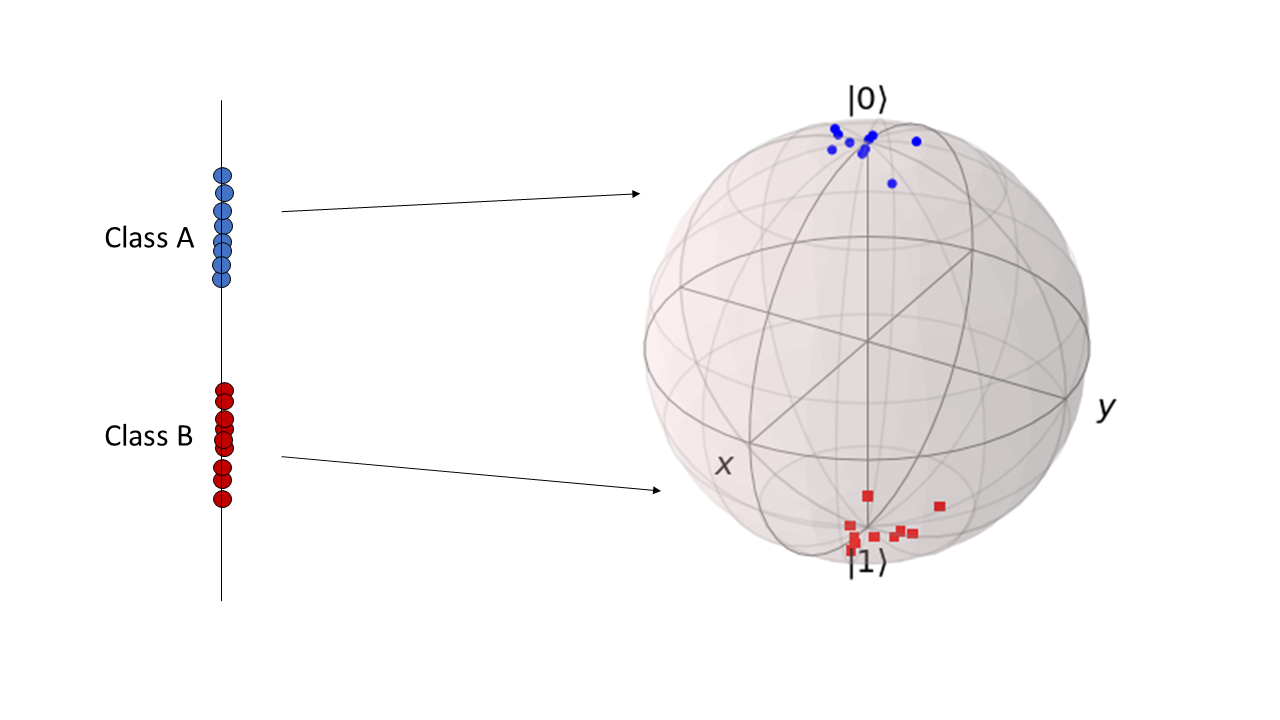}
    \caption{ {\bfseries Illustration of binary classification}. After the training stage, data from A are mapped to blue points, surrounding $\ket{0}$. Data from B are mapped to red points, surrounding $\ket{1}$. }
    \label{fig:Block}
\end{figure}
\begin{figure}[t]
    \includegraphics[scale = 0.45]{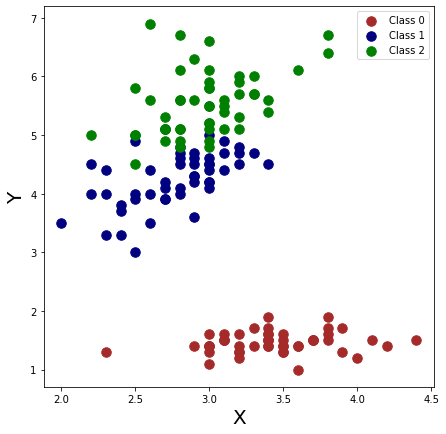}
    \caption{ The Iris Dataset. There are three classes distributed in a two-dimensional region. Different classes are represented by different colors. This dataset is used in our implicit approach for classification.}
    \label{fig: iris}
\end{figure}
\subsection{Explicit Approach}
\subsubsection{Construction}
\label{sec: explicit}
The implicit approach emphasizes training the embedding circuit to produce separated clusters on $\mathcal{H}$. However, the ``centers'' of these clusters are somewhat random. As long as relative distances among these clusters are maximal and those among data points within the same cluster are minimal, the method achieves its goal.

With the explicit approach, the cluster ``positions'' are designed to be fixed and separated into orthogonal subspaces. The main intuition is that with enough qubits, the Hilbert space is vast, complex, and can accommodate many smaller subspaces where data clusters can reside. These subspaces are well defined and well separated. If the data from different classes are ``approximately'' mapped to their proper subspaces, they are well separated by construction. The approximation here means that the embedded data might not exist completely within the desired subspace, instead possibly only in its vicinity. 
Then, we can ``measure'' the distance from a data point in $\mathcal{H}$ to different subspaces. Thus, classification of such a data point can be done accordingly.

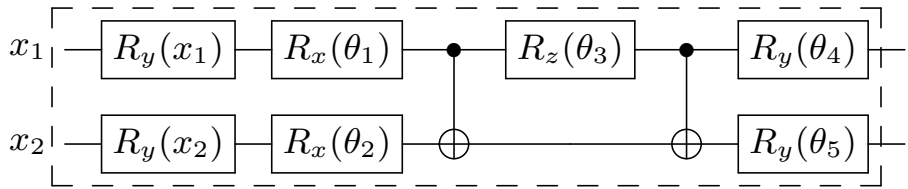
\begin{figure*}[t!]
        
        \scalebox{1.5}{
        \begin{minipage}{10cm}
        \Qcircuit @C=1em @R=1em {
        \lstick{x_1} & \gate{R_y(x_1)} & \gate{R_x(\theta_1)} & \ctrl{1} & \gate{R_z(\theta_3)} & \ctrl{1} & \gate{R_y(\theta_4)}  & \qw\\
        \lstick{x_2} & \gate{R_y(x_2)} & \gate{R_x(\theta_2)} & \targ    & \qw                  & \targ    & \gate{R_y(\theta_5)}  & \qw  \gategroup{1}{1}{2}{7}{.7em}{--} }
        \end{minipage}
        }
        
        \caption{{\bfseries Unit Embedding Circuit $\Phi$}. In implementation, this unit is repeated $4$ times. Hence, the total number of trainable parameters is $20$. In the end, the feature layer is repeated once more. The repetition of both feature and parameter layer has been used in Ref.~\cite{perez2020data} as a data re-uploading strategy that yielded better classification ability.}
    \label{fig: QAOA}
    \vspace{3mm}
\end{figure*}

Again, consider the supervised learning problem with $L$ labels, we decompose $\mathcal{H}$ into: 
\begin{equation}
\label{eqn: decomposition}
    \mathcal{H} = H_0 \oplus H_1 \oplus  ... \oplus H_{L-1} \oplus ..., 
\end{equation}
assuming that $L$ $\leq$ dim($\mathcal{H}$). We can always achieve this condition by adding more qubits to the circuits.

Our aim is to approximately map a data point accordingly to its ``label subspace'' $\{H_i\}$. Without loss of generality, let ${\rm dim}(H_i) = k$ and $H_i$ be spanned by $\{\ket{\psi_i^j}_{j=1}^k  \}$. Let a set of operators associated to label $i$, or equivalently, the subspace $H_i$, be: $\{    \ket{\psi_i^j}\bra{\psi_i^j} \}_{j=1}^k $. The likelihood of given data $\ket{x}$ having some label $i$ may be quantified by the projection of $\ket{x}$ onto the corresponding ``label'' subspace. Therefore, $\mathcal{M}_i$ could have the form: \\
\begin{equation}
    \mathcal{M}_i = \sum_{j=1}^k \ket{\psi_i^j}\bra{\psi_i^j}, 
    \end{equation}
which essentially is the targeted ensemble density operator (up to a normalization) associated with label $i$. 
The same strategy is then followed as we minimize the cost in Eq.~(\ref{eqn: cost}) and assign some unseen data $x$ according to the value of $f_i$. 


For example, we consider the binary supervised learning problem with $L = 2$ and 1-dim dataset $X = X_A \cup X_B$, where $X_A$ and $X_B$ are the training sets with label 0 and 1, respectively, as depicted in Fig.~\ref{fig:Block}. For simplicity, we use one qubit in the embedding circuit (refer to the 1-qubit toy model in \cite{lloyd2020quantum}). Hence, dim $\mathcal{H} = 2$. We then make the decomposition: $$   \mathcal{H} = \mathcal{H}_0 \oplus \mathcal{H}_1,    $$
where $\mathcal{H}_0$ and $\mathcal{H}_1$ are spanned by $\ket{0}$ and $\ket{1}$, respectively. We note the classifying vector $\vec{f}$ has the form:
\begin{equation}
\label{eqn: metricf}    
\begin{bmatrix}
\Tr{(\ket{x}\bra{x}\sigma_0)} \\
\Tr{(\ket{x}\bra{x}\sigma_1)} \\
\end{bmatrix}
= \begin{bmatrix}
\Tr{(\ket{x}\bra{x}\cdot\ket{0}\bra{0}) } \\
\Tr{(\ket{x}\bra{x}\cdot\ket{1}\bra{1}  )} \\
\end{bmatrix}.
\end{equation}

Following the same procedure helps obtain the cost value 
\begin{equation}
\label{eqn: costexplicit}
C = 1 - \frac{1}{2} {\rm Tr}[\sigma_z (\rho_A-\rho_B)],
\end{equation}
where $\rho_A$ and $\rho_B$ are state ensembles of the two respective training sets A and B and $\sigma_z$ is the Pauli-Z matrix. Minimization of this cost $C$ with respect to the circuit's parameters will give an embedding $\Phi(x,\theta)$ that maps the data from A to the vicinity of $\ket{0}$ in $\mathcal{H}$ and the data from B to the vicinity of $\ket{1}$ in $\mathcal{H}$, as illustrated in Fig.~\ref{fig:Block}.

An alternative picture also can be drawn from the described 1-dim dataset. If we choose the label space to be $\{H_0,H_1\}$, then the classifying vector in Eq.~(\ref{eqn: metricf}) can be obtained by simply performing measurement on the embedded state $\ket{x}$ in the $z$ basis.
Choosing a different label space, e.g., by decomposing $\mathcal{H} = \mathcal{H_+} \oplus \mathcal{H_-}$, where $\mathcal{H_+}$ and $\mathcal{H_-}$ are spanned by $({\ket{0}+\ket{1}})/{\sqrt{2}}$ and $({\ket{0}-\ket{1}})/{\sqrt{2}}$, respectively, measurement in the $x$ basis (i.e., the observable $\sigma_x$)  would need to be used instead. The classifying vector $\vec{f}$ in this latter case is a direct result from such a $\sigma_x$ measurement. Instead of the north and south pole on the Bloch sphere (Fig.~\ref{fig:Block}), the quantum circuit training would then make data points clustering around ``$+\hat{x}$ pole'' and ``$-\hat{x}$ pole,'' where $\hat{x}$ is the unit vector point along the positive $x$ direction.

\subsubsection{Connection to ``traditional'' QSL models }
By closely examining $\vec{f}$ in Eq.~(\ref{eqn: metricf}), the value of $f_0 = \Tr(\ket{x}\bra{x}\cdot\ket{0}\bra{0} ) = |\bra{0}\cdot\ket{x}|^2$ turns out to be the probability of obtaining state $\ket{0}$ when measuring the state $\ket{x}$ in computational basis. Most current quantum classifiers \cite{havlivcek2019supervised,schuld2020circuit} rely on these measurement outcomes after applying a general circuit $\Phi(x,\theta) = W(\theta)U(x)$ to some initial state $\ket{0}$ for classification. Hence, under the view of embeddings, by choosing the appropriate label space (specifically, the standard computational basis state), there is an intrinsic connection between this explicit approach and other traditional models~\cite{mitarai2018quantum, schuld2020circuit, schuld2019quantum}. More precisely, ``traditional'' approaches can be unified by this explicit approach. 

Such unification offers a two-fold advantage: the evaluation of overlaps between the input data $\ket{x}$ and $\ket{0}$ or $\ket{1}$, as in  Eq.~(\ref{eqn: metricf}), can be done simply by letting $x$ undergo the embedding circuit $\Phi(x,\theta)$ once and performing measurements instead of invoking the embedding circuit twice (in the SWAP test subroutine). Additionally, the cost evaluation in Eq.~(\ref{eqn: costexplicit}) does not necessarily need to be done in an iterative manner. In Refs.~\cite{adhikary2020entanglement, cao2020cost}, the authors provide elegant and efficient methods to encode the cost evaluation directly into quantum circuits. Hence, the training time can be reduced.  

\section{Numerical Simulations and Real-device Experiments}
With each approach, we train on the ideal simulator then use the optimized circuit to test on the ideal simulator, noisy simulator, and several real devices.

\subsection{Implicit Approach experiment}
\textit{Datasets}: For illustration purposes, we target the Iris dataset \cite{fisher1936use, anderson1936species} with $L = 3$ labels (Fig.~\ref{fig: iris}). There are $50$ data points in each class for a total of $N = 150$ data points. Ten data points are taken from each class to serve as the training data, and the remaining $40$ are used for testing. 
Aside from classification, our aim is to demonstrate the formation of clusters in the featured Hilbert space $\mathcal{H}$.

\textit{Quantum Embeddings}: We use the same so-called \textit{QAOA-like ansatz} as in~\cite{lloyd2020quantum} (Fig.~\ref{fig: QAOA}) for embedding.  The unit circuit $\Phi(x,\theta)$ is composed of a feature layer $U(x)$ followed by the parameter layer $W(\theta)$. Hence, we have $\Phi(x,\theta) = W(\theta)U(x)$, and the model is compact. A possible useful design of this embedding unit is to mix the feature parameters $x$ and tunable parameters $\theta$ to reduce the depth while maintaining the efficiency (see \cite{perez2020data}). For instance, instead of $R_y(x_1)$, one can consider $R_y( \theta_1, x_1  )$ or, generally, $R_y( g(\theta_1, x_1))$, where $g$ is some  function.

\begin{figure}[t]
    \includegraphics[scale=0.3]{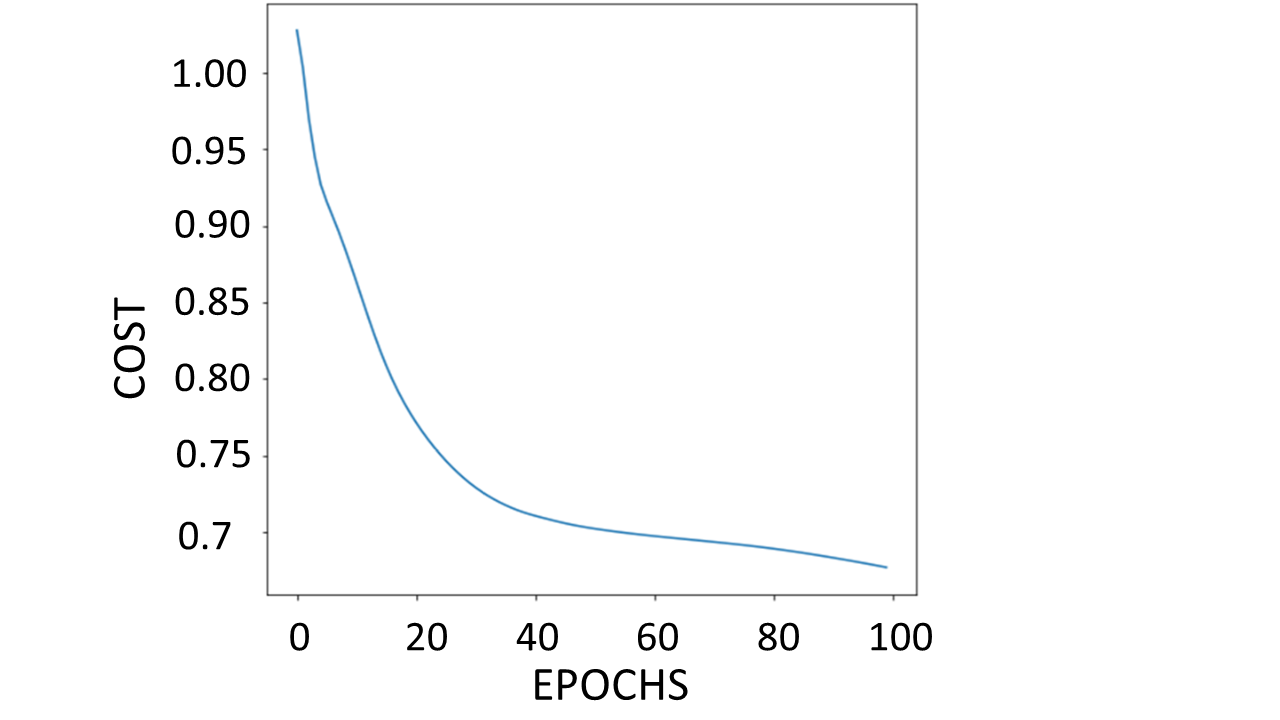}
    \caption{{\bfseries Cost as Function of Epochs}. In this training, we use $100$ epochs. There are $10$ training points for each class, totaling $30$ training points. }
    \label{fig: cost}
\end{figure}

\begin{figure}[h]
    \includegraphics[scale = 0.4]{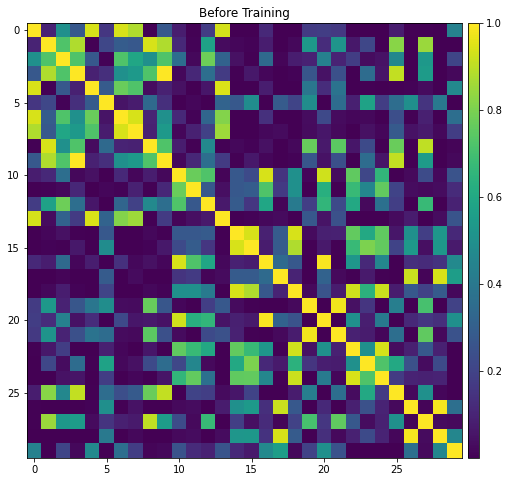}
    \includegraphics[scale =  0.4]{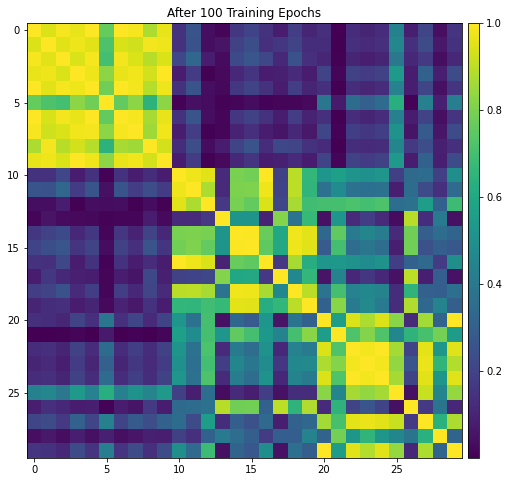}
    \caption{ \textbf{Visualization of overlaps between training points ($10$ training points in each class)}. (a) Top panel: Initial distribution of data in $\mathcal{H}$, in which the parameters in the variational quantum circuit are randomized. (b) Bottom panel: After $100$ training epochs, the data from the same class form a cluster on $\mathcal{H}$ as overlaps between their quantum states are high (brighter color). The visualization clearly shows that class $0$ (red points) are more separated from the other two classes. Meanwhile, classes $1$ (blue points) and $2$ (green points) are less separated from each other. The observation is in agreement with the testing results because all testing points from class $0$ are predicted with absolute accuracy, and false predictions mainly come from classes $1$ and $2$.}
    \label{fig: overlaps}
\end{figure}
\textit{Training Stage}: Because there are $N = 3$ labels in our problem, the cost function is:
\begin{equation}
   \label{eqn: cost3}
C = 1 - \frac{1}{3}\sum_{i=1}^3\Tr(\sigma_i)^2 + \frac{2}{3}\sum_{i<j}\Tr(\sigma_i\sigma_j).
\end{equation}
The training procedure is as follows:
\begin{itemize}
    \item Data from the training set are mapped to quantum states.
    \item Define and use SWAP test subprogram to evaluate $\Tr(\sigma_i^2)$ and $\Tr(\sigma_i\sigma_j)$. Later, we also use the inversion test.
    \item Minimize the cost $C$ in Eq.~(\ref{eqn: cost3}) over circuit parameters. 
\end{itemize}

Our simulation uses the PennyLane software package \cite{bergholm2018pennylane} and the optimization of $C$ over circuit parameters is done using the RMSprop \cite{Tieleman2012} optimizer with a learning rate of $0.01$.

\begin{table*}[t]
   \begin{tabular}{|c|c|c|c|c|c|c|c|}
\hline
     & U1 gate error & U2 gate error & U3 gate error & Readout error & CNOT error & Iris & Circles  \\
    \hline
{\tt ibmq\_16\_melbourne} & 0.0 & 0.00115 & 0.00229 & 0.06597 & 0.03157 & 92.5\% & 97\% \\
\hline
{\tt ibmq5\_yorktown} & 0.0 & 0.00084 & 0.00168 & 0.03494 & 0.02024 & 90.83\% & 93\%\\
\hline
{\tt ibmq\_bogota} & 0.0 & 0.00031 & 0.00062 & 0.03702 & 0.01171 & 92.5\% & 96\% \\
\hline
{\tt ibmq\_rome} & 0.0 & 0.00035 & 0.00071 & 0.02397 & 0.01344 & 92.5\% & 95\% \\
\hline

\end{tabular}    
    \caption{\textbf{Overall noise properties of four machines in our experiment.} For each column (e.g., ``U1 gate error''), we average over all the values of each qubit (i.e., U1 gate error of all qubits in the machine). For ``CNOT error rate,'' we average over all pairs of qubits. The last two columns show the results of noisy simulations.  }
    \label{tab: noiseproperties}
\end{table*}
\subsubsection{Results of noiseless simulations}
Figure~\ref{fig: cost} shows the training curve. As minimization takes place, the embedded data  in $\mathcal{H}$ are expected to form clusters, while those from different classes separate from each other. This is confirmed, as shown in Fig.~\ref{fig: overlaps}, in the comparison of the overlap of embedded data before and after the training. In particular, the overlaps between the embedded data from different classes become small after the training. This is especially the case for the overlap of class 0  with both class 1 and class 2. Thus, we have verified the well-separation of the embedded data from different classes.

After obtaining the optimized circuit parameters, we use the optimal circuit to perform a test on classifying the remaining unused data (i.e., the test dataset). The overall accuracy obtained is 92.5\%. Notably, only 30 data points (10 for each class) are used as the training set, which corresponds to 20\% of the total data points. This demonstrates that the classifier can classify unseen data with high accuracy, despite being trained with a relatively small training dataset (more details in Sec.~\ref{sec:smallsamples}). 



\subsubsection{Testing results from noisy simulations}
In real quantum hardware, noise and errors are important factors that reduce accuracy. To examine our method in the presence of noise, we test our classification with noisy models acquired from IBM Q ``backends.'' The device noise model is generated from their device calibration and accounts for gate error probability, gate length, $T_1$ and $T_2$ relaxation, and dephasing times, as well as the readout error probability. For convenience, Table~\ref{tab: noiseproperties} shows the average gate errors for the four backends considered in this work.

We test the classification with the SWAP test circuit via the noisy simulations, and the results are tabulated in Table~\ref{tab: summaryofresults}. The accuracy seems to be unaffected by the noise, and the values from the noisy simulator using the four noisy models from the respective devices are 92.5\%, 90.83\%, 92.5\%, and 92.5\%. The circuit parameters used are obtained from the noiseless optimization. The reason for not using noisy simulators to obtain the parameters is because we will perform the same testing on the actual hardware. Hence, it would be impractical and too time consuming to perform the training directly on the hardware as the jobs queue could be long and execution of the training circuits would have to be split over many jobs.

\subsubsection{Run on quantum computers}
With the noisy simulation, we also test our ideally trained model on real quantum ``backends.'' Table~\ref{tab: summaryofresults} summarizes the detailed results, including simulations and real devices. For the same classification, we run two different methods to obtain overlaps: the SWAP test, which uses five qubits, and the inversion test that uses only two qubits. 

\begin{figure*}
    \centering
    \includegraphics[width = 1.0\linewidth]{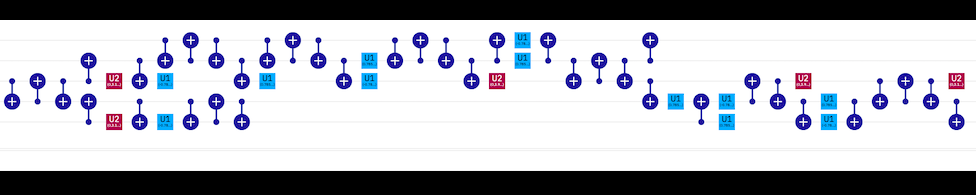}
    \caption{ \textbf{Decomposition of controlled-SWAP gates into many CNOT and one-qubit gates.} In our work,  classical data are embedded in 2-qubit states. Hence, there are five total qubits in the SWAP test circuit that uses a controlled-SWAP gate. \textit{Note}: this diagram only shows the decomposition of c-SWAP, not including the data embeddings part. There are 38 CNOT gates used in the decomposition.  }
    \label{fig: cnotdecomopse}
\end{figure*}

Accuracy with the SWAP test ranges from 28\% to 75\% on various backends. However, inversion test accuracy remains stable around 90\%. This clearly shows substantial performance differences between the SWAP and inversion tests. Likely, the main factor that accounts for such discrepancy is the controlled SWAP gate (c-SWAP in Fig.~\ref{fig: SwapandInver}) in the SWAP test circuit. The number of c-SWAP gates required for two $n$-qubit states scales as $ \mathcal{O}(n)$. Each c-SWAP gate then is decomposed into many CNOT gates (which are noisy) as shown in Fig.~\ref{fig: cnotdecomopse}. Despite the noisy simulations yielding accuracy around 90\%, runs on the actual machines suffer accumulated errors not captured in the noise model used in the simulations.
On the other hand, the inversion test does not need the c-SWAP gate and, hence, requires fewer CNOTs---but at the cost of doubling the quantum circuit depth. In our classification model, there is a trade-off between using either the SWAP or inversion test. As our small-size experiments have shown, the inversion test should be used for better classification on NISQ machines. However, it requires the ability to invert the embedding circuit and can only evaluate the overlaps between two data points (of pure states). Hence, classifying unseen data (by obtaining the classifying vector $\vec{f}$) must be done in an iterative manner. Conversely, the SWAP test can handle mixed states, and the classification can be sped up with a QRAM. In addition, SWAP test performance can be further improved by using methods introduced in~\cite{cincio2018learning}. Such an approach is hardware-dependent (as the authors examined on IBM Q and Rigetti separately), and it requires fewer CNOT gates.\footnote{A similar discussion regarding the inversion test and the method in~\cite{cincio2018learning} also is presented in~\cite{havlivcek2019supervised}. Our experimental work elaborates on this issue further as we have explicitly examined the SWAP and inversion tests' performance in the presence of noise.}

\begin{table}
    \centering
    \begin{tabular}{|c|c|c|}
\hline
     & Iris & Circles  \\
    \hline
\textit{Ideal simulator} & 92.5\% & 96\% \\ 
\hline
{\tt ibmq\_16\_melbourne} & 
\begin{tabular}{c}92.5\% (n.s.) \\
32.5\% (r.d. ST) \\ 88.33\% (r.d. IT) 
\end{tabular}  & 
\begin{tabular}{c}97\% (n.s.) \\
99\% (r.d.)
\end{tabular}\\
\hline
{\tt ibmq\_5\_yorktown} &
\begin{tabular}{c}90.83\% (n.s.)\\
50\% (r.d. ST) \\
83.83\% (r.d. IT) 
\end{tabular}& 
\begin{tabular}{c}93\% (n.s.) \\
91\% (r.d.)
\end{tabular}
\\
\hline
{\tt ibmq\_bogota} & \begin{tabular}{c}92.5\%  (n.s.) \\ 75\%  (r.d. ST) \\ 91.67\% (r.d. IT) \end{tabular} & \begin{tabular}{c}96\% {\rm (n.s.)} \\ 95\% {\rm (r.d.)}\end{tabular} \\ \hline
{\tt ibmq\_rome} &  \begin{tabular}{c}92.5\% {\rm (n.s.)} \\ 28.33\% {\rm (r.d. ST)} \\ 91.67\% {\rm (r.d. IT) }\end{tabular}   & \begin{tabular}{c}95\% {\rm (n.s.)} \\ 94\% {\rm (r.d.)}\end{tabular} \\
\hline
\end{tabular}    
    \caption{\textbf{Summary of results.} For simulations and runs on actual backends for both the Iris and make\_circles datasets.
    The result on the top of each block corresponds to noisy simulation (n.s.), and the bottom ones correspond to real device (r.d.). ``ST'' denotes SWAP Test, and ``IT'' represents Inversion Test.  }
    \label{tab: summaryofresults}
\end{table}

\subsection{Explicit Approach experiment}

\begin{figure}[htbp]
    \includegraphics[scale = 0.36]{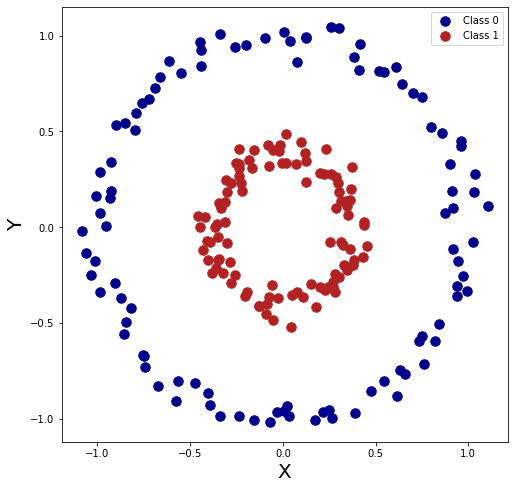}
    \caption{The Make\_circles dataset with $N = 2$ labels used in our explicit approach for classification. }
    \label{fig: makecircles}
\end{figure}

\textit{Datasets}: To illustrate this approach, we target the dataset {\tt make\_circles} with $N = 2$ labels (Fig. \ref{fig: makecircles}). 
Fifteen points are taken from each class to serve as the training data. For testing, we generate an additional $50$ points for each class.

\smallskip\textit{Training Stage}: We choose two subspaces spanned by $\ket{00}$ and $\ket{11}$, respectively, as label spaces and train the circuit to map data from class $0$ (blue points) to $H_{00}$ and from class $1$ (red points) to $H_{11}$. Given some input data $x$, the classifying vector is then:
\begin{equation}
\label{eqn: metricexplicit}
\begin{bmatrix}
\Tr{(\ket{x}\bra{x}\cdot\ket{00}\bra{00}) } \\
\Tr{(\ket{x}\bra{x}\cdot\ket{11}\bra{11})  } \\
\end{bmatrix}.
\end{equation}
The cost function becomes:
\begin{eqnarray}
C& =& 1-\frac{1}{2}\Big\{\Tr{\big[\sigma_A(\ket{00}\bra{00}-\ket{11}\bra{11})\big]} \nonumber\\
&&- \Tr{\big[\sigma_B(\ket{00}\bra{00}-\ket{11}\bra{11})\big]}\Big\},     \end{eqnarray}
where $\sigma_A = \frac{1}{N_A} \sum_A \ket{x_A}\bra{x_A}$ and $\sigma_B = \frac{1}{N_B} \sum_B \ket{x_B}\bra{x_B}$. A and B refer to class 0 and class 1, respectively. 


\subsubsection{Results of noiseless simulations}
\begin{figure}
    \includegraphics[scale = 0.3]{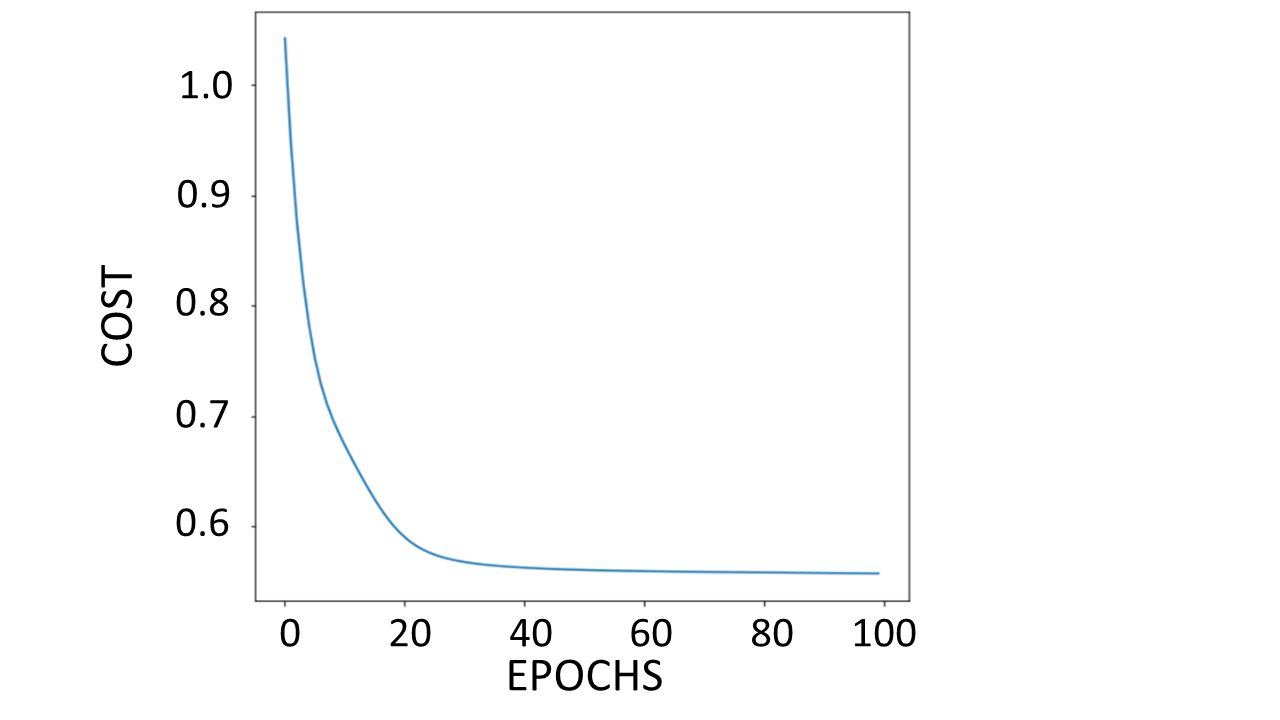}
    \caption{ \textbf{Cost as a function of epochs.} There are $100$ epochs in this training with $15$ training points for each class and $30$ total training points. }
    \label{fig: costexplicit}
\end{figure}

\begin{figure}
    \includegraphics[scale  = 0.45]{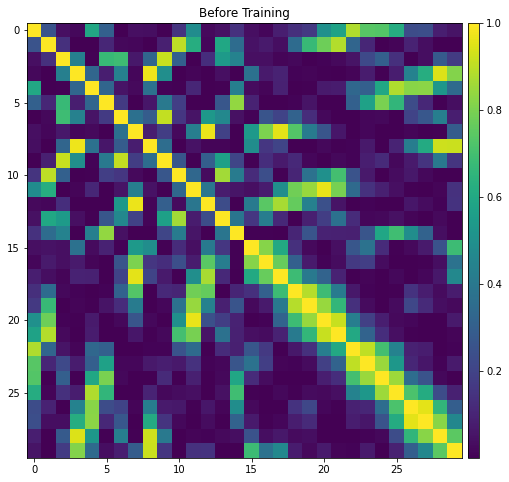}
    \includegraphics[scale = 0.45]{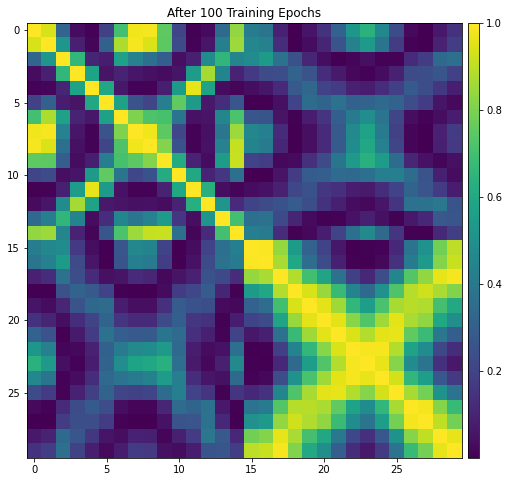}
    \caption{\textbf{Visualization of overlaps between training points (15 training points in each class)}. After training process, data from different classes become separated. Compared with the implicit approach, the ``well-separation'' is less apparent. In other words, clusters are less tight. This may be reasonably explained by the way two methods work. In the implicit approach, the optimization procedure focuses on \textit{directly} separating data points from different classes. Meanwhile in the explicit approach, data get separated \textit{indirectly}, i.e, through the pre-defined subspaces. Hence, in theory, the implicit approach has certain advantages over the explicit method in terms of attaining complete separation. In practice, both approaches have strong classification ability, demonstrated in our experiments.  }
    \label{fig: visualizaexplicit}
\end{figure}

Figure~\ref{fig: costexplicit} presents the training curve in the noiseless simulation. The overall accuracy is 96\%. As in the implicit case, we visualize the overlaps between training points before and after training with 100 epochs (Fig.~\ref{fig: visualizaexplicit}). This confirms the well-separation of embedded data from different classes in our explicit approach.

\subsubsection{Noisy simulations}
As in the previous implicit case, we also use the trained parameters from the noiseless simulation for the quantum embedding but perform the noisy simulation to classify the testing dataset. The accuracy of these noisy simulations is tabulated in Table~\ref{tab: summaryofresults}. Noise details can be found in Table~\ref{tab: noiseproperties}.


\subsubsection{Run on quantum computers}
We perform the classification experiments of this explicit approach on real quantum backends {\tt ibmq\_16\_melbourne}, {\tt ibmq\_5\_yorktown}, {\tt ibmq\_bogota}, and {\tt ibmq\_rome} and compare their accuracy to the noisy simulation with the noise model from the same backend. Table~\ref{tab: summaryofresults} summarizes the results.

The accuracy from the noisy simulators and real machines turns out to agree well with each other, achieving values above 90\%. This indicates that the explicit approach is less affected by the noise compared to the implicit approach using the SWAP test. This is reasonable as the explicit approach requires less resources, such as fewer two-qubit gates, than the implicit approach with the SWAP test. We simply let the input data run through the circuit and perform computational-basis measurements as the classification only depends on the probability of obtaining $\ket{00}$ and $\ket{11}$. 


\subsection{Training over small samples }\label{sec:smallsamples}
\begin{figure}[htbp]
    \includegraphics[scale = 0.5]{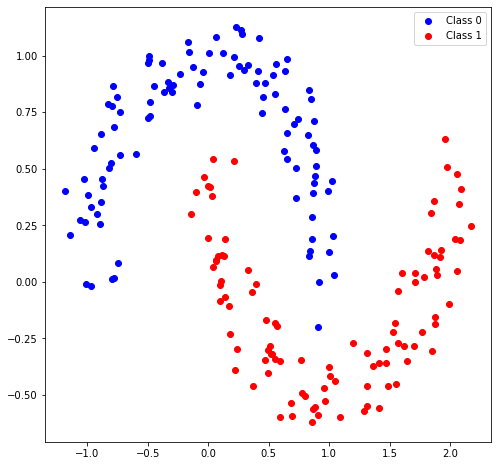}
    \caption{The make\_moons dataset.  }
    \label{fig: moon}
\end{figure}

We observe that both approaches produce surprisingly good testing accuracy despite training with small data points. This provides motivation to determine whether or not one approach is more robust than the other in terms of learning capacity. Another motivator is to investigate how testing accuracy varies with training size. We choose the {\tt make\_moons} dataset with $L = 2$ labels to perform the numerical experiment (Fig.~\ref{fig: moon}). 

The procedure is as follows: 
\begin{itemize}
    \item For each class, we generate a fixed set of 50 points (hence, there are 100 points in total), serving as testing instances.
    \item For each class, we then choose randomly 5, 10, 20, and 25 points, serving as training instances. For each number of training instances, we average over multiple training to obtain testing accuracy.
    \item We train the quantum circuit using implicit and explicit approaches separately and compare the testing accuracy after 100 epochs of training.
\end{itemize}
Table~\ref{ tab: smallsizetesting } summarizes the results. \\
\begin{table}
    \centering
    \begin{tabular}{|c|c|c|}
    \hline
    Training size  & Implicit Approach & Explicit Approach  \\
    \hline
    5   & 84\%    & 75.8\% \\
    7   & 84\%    & 80.4\%  \\
    10  & 83.4\%  & 80.4 \% \\
    15  & 87\%   & 86\% \\
    20  & 87.5\%    & 86.5 \% \\
    25 & 89.5\%  &  92\% \\
    \hline
    \end{tabular}
    \caption{\textbf{Summary of testing results from make\_moons dataset.} There are 100 points in each class for a total of 200. For each class, we choose 50 points to serve as testing instances. The remaining 50 are used as the pool to randomly select a certain size of training instances as described previously.}
    \label{ tab: smallsizetesting }
\end{table}

Evidently, the implicit approach performs well---even with a small training size---which implies a certain advantage in this approach. As the training size increases, both approaches tend to achieve equal performance.

\section{Prospect in the NISQ era}
\label{sec: prospect}

\begin{figure}[ht]
    \includegraphics[width = 1.\linewidth]{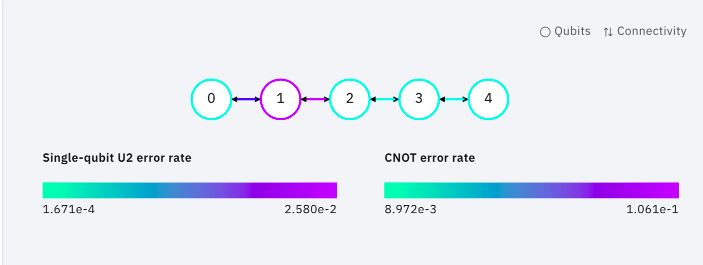}
    \caption{\textbf{Topology of {\tt ibmq\_bogota}}. The picture was acquired at the time of our experiment. Topologies of other machines used in our experiment can be found in Fig.~\ref{fig: devicetopology}. }
    \label{fig: bogotatopology}
\end{figure}

To maximally enhance the performance of any quantum algorithm or, generally, a quantum procedure, we also need to take into account the hardware structure, e.g., the connectivity of qubits in the system and specific qubits chosen. Figure~\ref{fig: bogotatopology} shows the topology of the machine {\tt ibmq\_bogota} used in our work. 

 Table~\ref{tab: summaryofresults} shows the result of testing the Iris dataset on {\tt ibmq\_bogota} with the inversion test done using qubits labeled ``3'' and ``4.'' We also carried out the same ``testification'' using qubits ``1'' and ``2.'' The testification result 65.83\% is dramatically lower than using qubits 3 and 4 (91.67\%). Such deviation can be reasonably argued from the noise rates of the qubit pair involved and their CNOT gates. As depicted in Fig.~\ref{fig: bogotatopology}, qubits 3 and 4, as well as the connection between them, have much lower error rates compared to those of qubits 1 and 2. Hence, in practice, any quantum procedure needs to be hardware-aware to reach its maximum efficiency. Of course, our experiment requires very few numbers of qubits and simple gates. As such, we can simply choose specific qubits to obtain better accuracy. More complicated quantum circuits generally require more careful qubit specification. The quantum hardware topology also varies, e.g., {\tt ibmq\_16\_melbourne} and {\tt ibmq5\_yorktown} versus {\tt ibmq\_bogota}. Such differences can affect the decomposition of a multi-qubit gate into available one- and two-qubit (in particular) gates. A hardware-aware compiler that optimizes the selection also is a necessity for future large-scale tasks.
This hardware-specification optimization is important practically and requires additional development. Given an arbitrary quantum backend's topology and description of some quantum procedures, such as a circuit's length, width, and number of 1-qubit or 2-qubit gates, it may be worth determining if a systematic procedure exists that can decide which qubits---and in what orders---should be used in order to maximize the performance. 

Other works also have demonstrated the ability and feasibility of using actual quantum computers to classify real-world data~\cite{blank2020quantum, havlivcek2019supervised}. In addition to providing a unified framework for QSL, we have performed simulations and cloud-based real-device experiments. These experiments on real quantum backends have extended the prospect of applying quantum computers for ML one step further, demonstrated explicitly in our work showing current noisy quantum computers can achieve high accuracy on classifying data. The low-accuracy results obtained via the SWAP test routine may be improved by using the inversion test routine, and we have emphasized that the inversion test is more appropriate in the NISQ era. Real quantum systems undoubtedly are more complicated, and their noise on long circuits (especially those with many CNOTs) may result in worse accuracy than noisy simulations. In our experiments, the error rate of 1-qubit gates is $\sim 10^{-3}$ or smaller, and the 2-qubit CNOT is $\sim 10^{-2}$ for current hardware~\cite{preskill2018quantum}. Full implementation of quantum error correction remains elusive. Along with development of precise, high-fidelity gates, efforts have been made in error mitigation methods~\cite{temme2017error,endo2018practical,Chen2019DetectorTO,bravyi2020mitigating} to obtain useful outcomes. Some of these mitigation methods require repetition of the same circuit but with different overall error rates by possibly stretching the gate pulses. This allows observables to be extrapolated to the gate noiseless limit~\cite{temme2017error,endo2018practical}. Error mitigation measurement also is necessary to infer correct readout outcomes~\cite{Chen2019DetectorTO,bravyi2020mitigating}. 
The experiments done as part of our work do not employ any mitigation technique. As such, our results can be further improved with these techniques, especially results from the SWAP test using gate mitigation. Additionally, our classification model has been shown to do well with a small training pool (compared to the testing set), achieving very high accuracy. Hence, one can reasonably expect that the model can be trained on real machines to achieve comparable performance.

\begin{figure}[htbp]
    \includegraphics[width = 0.95\linewidth]{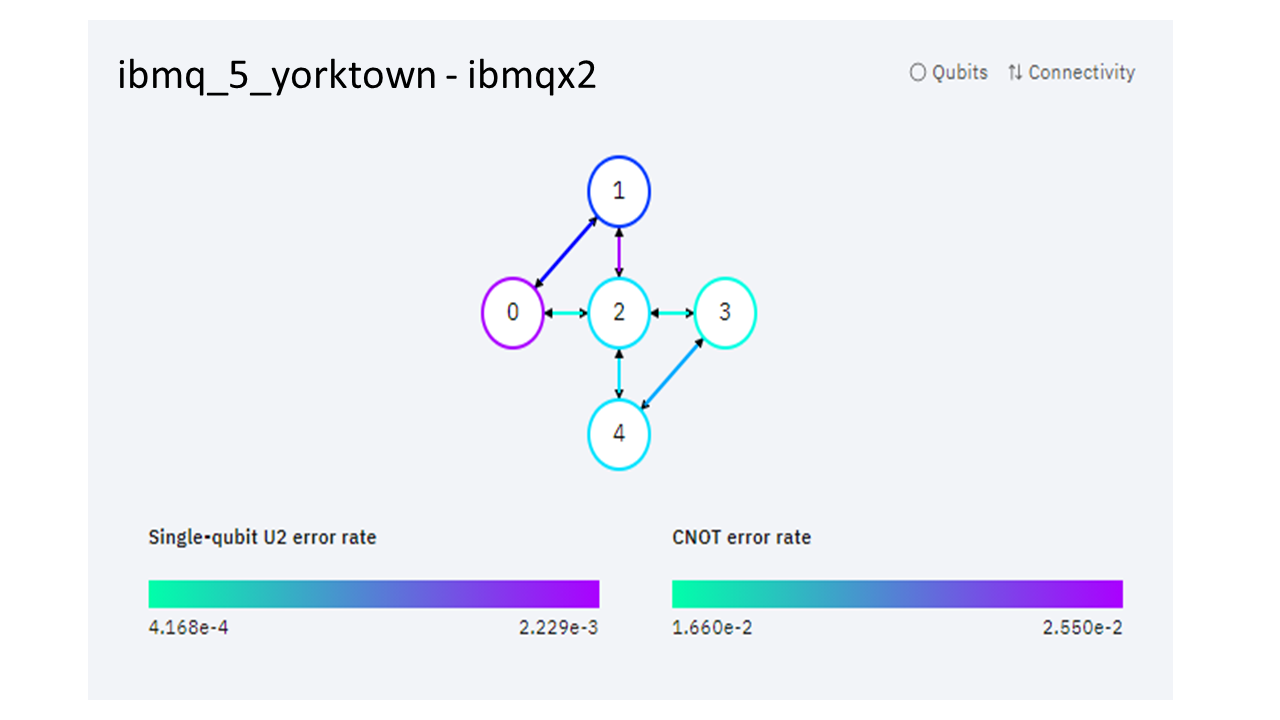}
    \includegraphics[width = 0.95\linewidth]{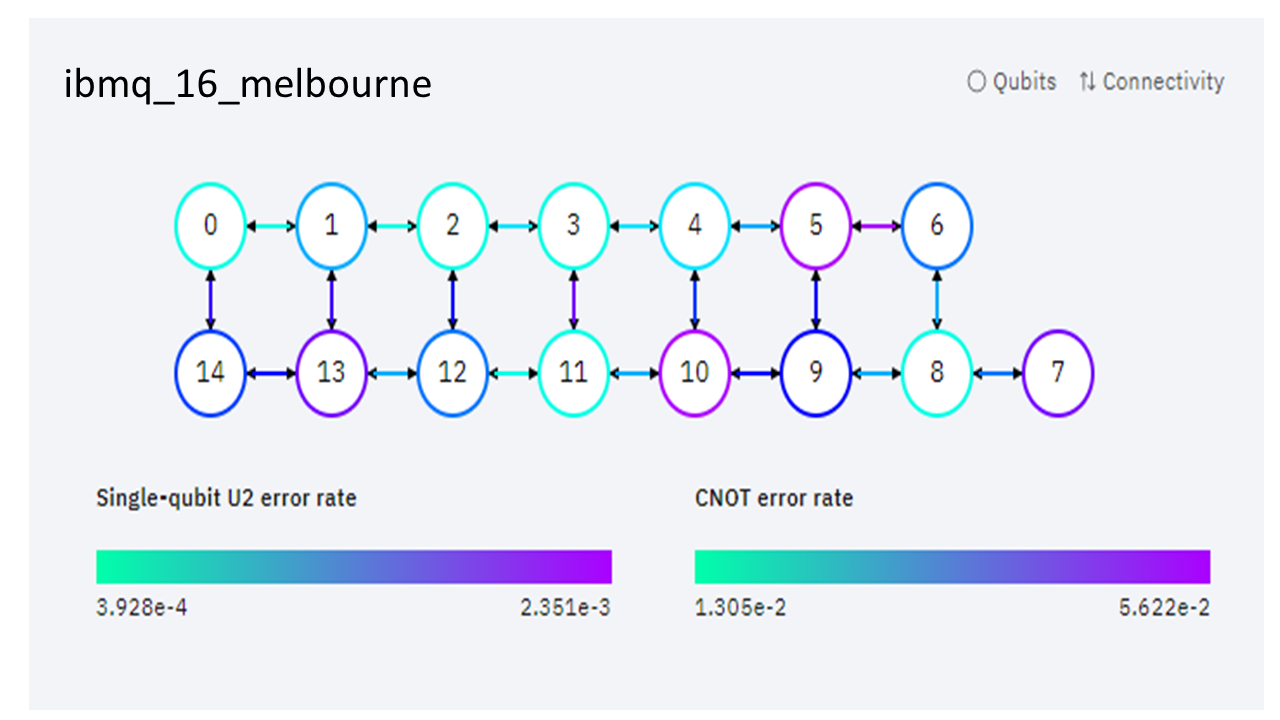}
    \includegraphics[width = 0.9\linewidth]{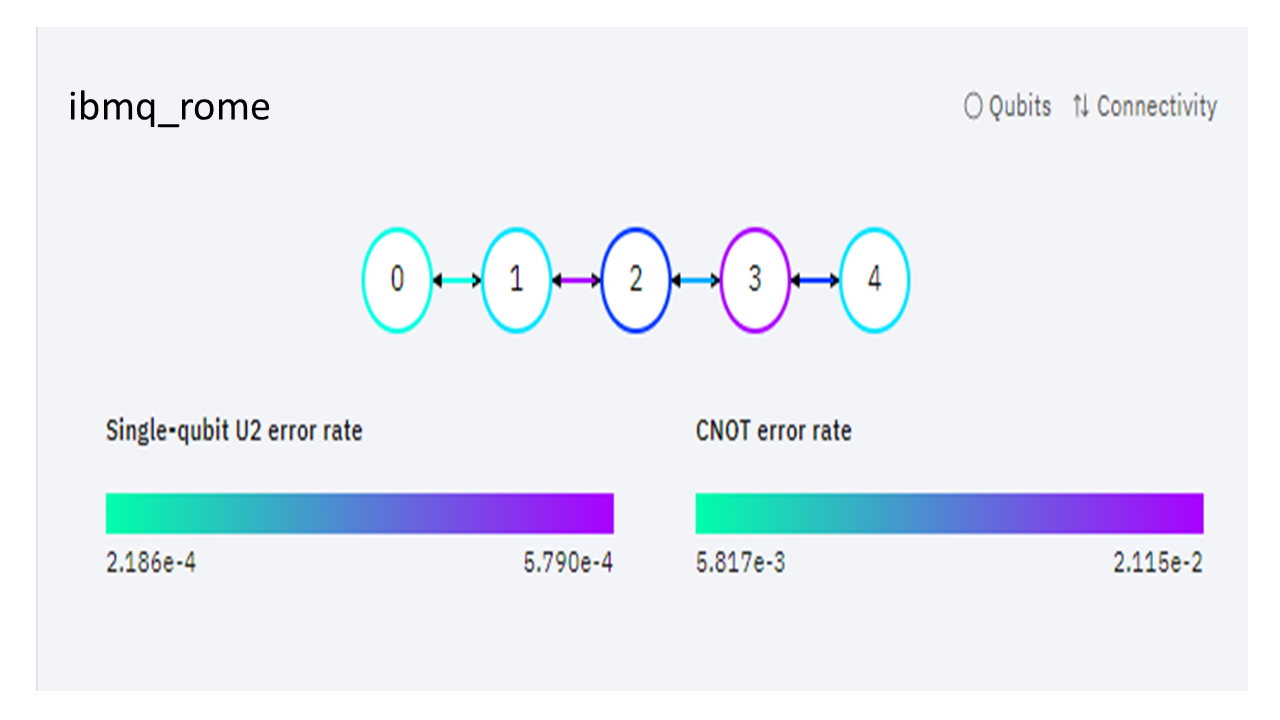}
    \caption{Topology and the coupling map of other IBM Q devices used in this work: {\tt ibmq\_5\_yorktown}, {\tt ibmq\_16\_melbourne}, and {\tt ibmq\_rome}. }
    \label{fig: devicetopology}
\end{figure}
\section{Conclusion}
\label{conclude}

In our framework for quantum supervised learning, the main conceptual tool of our method is the idea that the input data $x$ is ``forwarded'' to the classifying vector $\vec{f}$, and their classification can be done accordingly. A hybrid optimization step then proceeds to train the circuit. After being trained, the embedding circuit can map the data from the input space $\mathcal{X}$ to the proper subspaces in $\mathcal{H}$.

Our work emphasizes that the quantum feature map, equipped with a learning procedure, is an especially powerful tool for supervised learning. With the implicit approach, the number of separated classes (labels) in a supervised learning problem ideally can be arbitrarily high. Thus, it provides a means to construct a universal quantum classifier. Compared to the explicit approach, the learning capacity of this approach has been demonstrated with a small training pool, which is also encouraging. Moreover, we show that the explicit approach can intrinsically unify other \textit{traditional} QSL models (detailed in Appendix \ref{sec: unification}). The fact that our framework can be divided into explicit and implicit approaches demonstrates its flexibility, affording the option to choose how data are embedded and analyzed on the Hilbert space $\mathcal{H}$. These two approaches constitute a unified framework for supervised learning methods using a quantum computer. 

Along with classification, we note that the trained quantum circuit possibly can be employed as a subroutine of other quantum ML algorithms as we know with high confidence that embedded data from different classes are well separated from each other (trained with the implicit approach) or approximately well contained in some subspaces in $\mathcal{H}$ (trained with the explicit approach).


\section*{Acknowledgments}
We acknowledge use of the IBM Q for this work. The views expressed are those of the authors and do not reflect the official policy or position of IBM or the IBM Q team. This research used resources of the Oak Ridge Leadership Computing Facility, which is a U.S.\ Department of Energy, Office of Science (DOE-SC) User Facility supported under Contract DE-AC05-00OR22725. This work was partially supported by the DOE-SC Office of High Energy Physics program under Award Number DE-SC-0012704 (S.Y-.C.C.), Brookhaven National Laboratory LDRD \#20-024 (S.Y-.C.C.), and a subcontract No.~384153 from Brookhaven Science Associates LLC (T.-C.W.).

\appendix

\section{SWAP and Inversion tests}
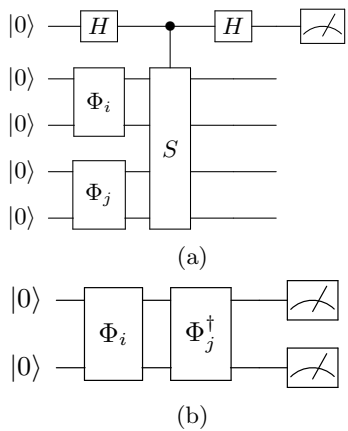
\begin{figure}[htbp]
     \begin{subfigure}[b]{0.3\textwidth}
         \centering
         \scalebox{1.0}{
            \Qcircuit @C=1em @R=1em {
            \lstick{\ket{0}} & \gate{H}  & \ctrl{1} &  \gate{H}  &\qw & \meter \\
            \lstick{\ket{0}}&  \multigate{1}{\Phi_i} & \multigate{3}{S} & \qw & \qw \\
            \lstick{\ket{0}} & \ghost{U_i} & \ghost{S} & \qw  & \qw\\
            \lstick{\ket{0}} &  \multigate{1}{\Phi_j} &  \ghost{S} & \qw  & \qw  \\
            \lstick{\ket{0}} & \ghost{U_j} & \ghost{S}  & \qw   & \qw  
            }
            
            }
            \caption{}
     \end{subfigure}
     \hfill
     \begin{subfigure}[b]{0.3\textwidth}
         \scalebox{1.15}{
            \Qcircuit @C=1em @R=1em {
            \lstick{\ket{0}} & \multigate{1}{\Phi_i}& \multigate{1}{\Phi_j^{\dagger}} & \qw & \meter \\
            \lstick{\ket{0}} & \ghost{U_i} & \ghost{U_j} & \qw & \meter \\
            }
            
            }
            \caption{}
     \end{subfigure}
        \caption{\textbf{Circuit representation for SWAP test (top) and inversion test (bottom).} There is an abuse of notation $\Phi$: in both the SWAP and inversion test circuits, the actual embeddings circuit $\Phi_{i,j}$ already includes the repetition of the unit embedding in Fig. \ref{fig: QAOA}.}
        \label{fig: SwapandInver}
\end{figure}

Figure~\ref{fig: SwapandInver} provides a description of two alternative methods to evaluate the overlaps between two data points. The first is the control-SWAP gate, acting on five qubits, and the second is the inversion test.



%

\section{More on the Explicit Approach}
\label{sec: unification}

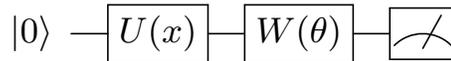
\begin{figure}[htbp]
\begin{center}
    \scalebox{1.5}{
    \Qcircuit @C = 1em @R = 1em {
    \lstick{\ket{0}} & \gate{U(x)} & \gate{W(\theta)}  & \meter \\  
    }
    }
    \caption{A common circuit model for machine learning tasks. }
    \label{fig: singlequbitcircuit}
\end{center}
\end{figure}

In Section \ref{sec: explicit}, we have illustrated the intrinsic relation between the explicit approach and ``traditional'' quantum supervised learning (QSL) models via the measurement outcomes. Here, we offer an alternative explanation. We consider a binary classification problem (two classes, A and B) and a single-qubit quantum circuit (Fig.~\ref{fig: singlequbitcircuit}). 

In traditional approaches, the first block $U(x)$ embeds classical information $x$ into a quantum state in the Hilbert space $\mathcal{H}$. Then, the variational layer $W(\theta)$ is trained to distinguish those embedded states. For instance, $x$ can be assigned to class A if the probability of measuring 0 is greater than the probability of measuring 1 ($P_0 > P_1$). The training step should focus on maximizing the probability of measuring 0 for data in class A and 1 otherwise. 

Recall that our embedding-based framework exploits the ability of a quantum circuit to represent data in a complex space. An alternatively simple perspective is evident if we interpret the whole circuit as an encoding of the classical data $x$.  
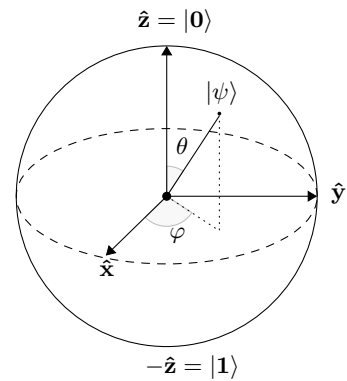
\begin{figure}
\begin{tikzpicture}[line cap=round, line join=round, >=Triangle]
  \clip(-2.19,-2.49) rectangle (2.66,2.58);
  \draw [shift={(0,0)}, lightgray, fill, fill opacity=0.1] (0,0) -- (56.7:0.4) arc (56.7:90.:0.4) -- cycle;
  \draw [shift={(0,0)}, lightgray, fill, fill opacity=0.1] (0,0) -- (-135.7:0.4) arc (-135.7:-33.2:0.4) -- cycle;
  \draw(0,0) circle (2cm);
  \draw [rotate around={0.:(0.,0.)},dash pattern=on 3pt off 3pt] (0,0) ellipse (2cm and 0.9cm);
  \draw (0,0)-- (0.70,1.07);
  \draw [->] (0,0) -- (0,2);
  \draw [->] (0,0) -- (-0.81,-0.79);
  \draw [->] (0,0) -- (2,0);
  \draw [dotted] (0.7,1)-- (0.7,-0.46);
  \draw [dotted] (0,0)-- (0.7,-0.46);
  \draw (-0.08,-0.3) node[anchor=north west] {$\varphi$};
  \draw (0.01,0.9) node[anchor=north west] {$\theta$};
  \draw (-1.01,-0.72) node[anchor=north west] {$\mathbf {\hat{x}}$};
  \draw (2.07,0.3) node[anchor=north west] {$\mathbf {\hat{y}}$};
  \draw (-0.5,2.6) node[anchor=north west] {$\mathbf {\hat{z}=|0\rangle}$};
  \draw (-0.4,-2) node[anchor=north west] {$-\mathbf {\hat{z}=|1\rangle}$};
  \draw (0.4,1.65) node[anchor=north west] {$|\psi\rangle$};
  \scriptsize
  \draw [fill] (0,0) circle (1.5pt);
  \draw [fill] (0.7,1.1) circle (0.5pt);
\end{tikzpicture}
\caption{\textbf{Visualization of a qubit on a Bloch sphere.} Note that the angle $\theta$ in this figure differs from the circuit parameters $\theta$ in Fig. \ref{fig: singlequbitcircuit}. }
    \label{fig: blockunification}
\end{figure}

Without loss of generality, we assume classical data $x$ are mapped to a quantum state $\ket{\psi}$ (refer to Fig.~\ref{fig: blockunification}). When we measure this state, the probability of measuring 0 is $\cos^2 (\theta)$, which means that if such a probability is high, the state $\ket{\psi}$ will be close to the ``north pole'' $\ket{0}$. If we follow the explicit approach and decompose $\mathcal{H} = \mathcal{H}_0 \oplus \mathcal{H}_1$, where $\mathcal{H}_0, \mathcal{H}_1$ are spanned by $\ket{0}, \ket{1}$, respectively, we end up maximizing the probability of measuring 0 for data from class A and measuring 1 for data from class B. Pictorially, those data from class A will form a cluster around $\ket{0}$, and data from class B will cluster around $\ket{1}$. After optimization, these clusters are ``well-separated.'' To classify unseen data, we can use the optimized circuit to map them to some states and measure. The measurement probability may be understood as a ``closeness'' to either one of the two data cluster ``centers'' from classes A and B. This example also illustrates that our embedding-based framework, or, more specifically, the explicit approach, conceptually unifies other \textit{traditional} QSL models.

\section{One-versus-all strategy}
\label{caveatovall}
\begin{figure}[htbp]
    \includegraphics[width = 1.0\linewidth]{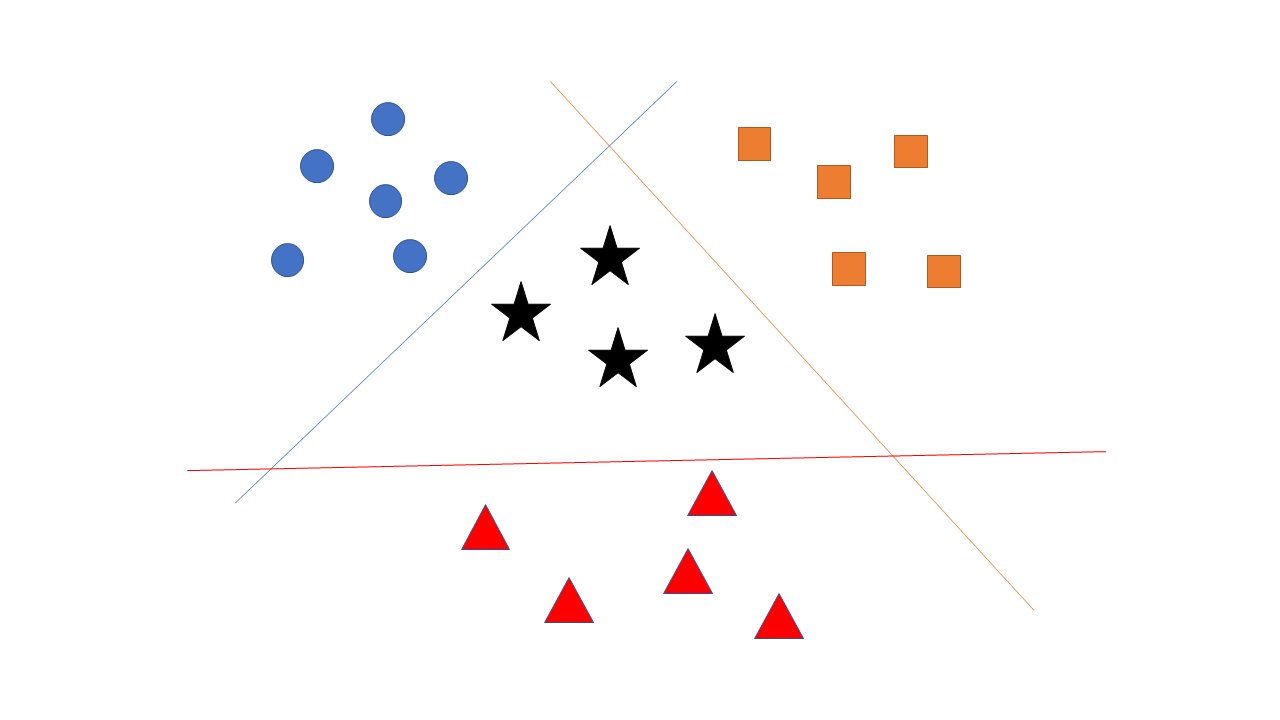}
    \caption{\textbf{One-versus-all Strategy.} There are three classes (blue, orange, and red) and the corresponding decision boundary (blue, orange, and red lines). All classes are linearly separable for simplicity. }
    \label{fig: onevall}
\end{figure}

The one-versus-all strategy (Fig.~\ref{fig: onevall}) often has been used to transform a binary classifer to a multi-class classifer, especially for those models that, in essence, can deal with only binary classification. Here, we review this strategy and discuss its drawbacks.

The underlying mechanism of the one-versus-all strategy is it assumes there are only two classes (or labels) in the supervised learning problem that learn the corresponding decision boundary. For example, in Fig.~\ref{fig: onevall}, all blue circles and red triangles can be treated as one class that learns the decision boundary to distinguish them from the orange rectangles, as well as the decision boundary between the blue circles and the rest and the red triangles from the others. \\

The caveat of the one-versus-all strategy is clear from Fig.~\ref{fig: onevall}: the black stars (unseen data) struggle to find a class (label). A  similar issue appears in QSL, where the data are embedded by a fixed circuit in the Hilbert space $\cal{H}$ and  the subsequent variational circuit is trained to draw the decision boundary. Notably, our framework can naturally surpass this issue as the representation of the data in $\cal{H}$ is learned. Then, a measure is employed to compare data directly as in the implicit approach or indirectly like the explicit approach. Hence, a label for unseen data is always guaranteed. \\

\bibliography{ref,bib/qml_examples}{}

\begin{thebibliography}{10}

\bibitem{nielsen2002quantum}
Michael~A Nielsen and Isaac Chuang.
\newblock Quantum computation and quantum information, 2002.

\bibitem{harrow2017quantum}
Aram~W Harrow and Ashley Montanaro.
\newblock Quantum computational supremacy.
\newblock {\em Nature}, 549(7671):203--209, 2017.

\bibitem{arute_quantum_2019}
Frank Arute, Kunal Arya, Ryan Babbush, Dave Bacon, Joseph~C Bardin, Rami
  Barends, Rupak Biswas, Sergio Boixo, Fernando~GSL Brandao, David~A Buell, and
  {others}.
\newblock Quantum supremacy using a programmable superconducting processor.
\newblock {\em Nature}, 574(7779):505--510, 2019.

\bibitem{briegel2009measurement}
Hans~J Briegel, David~E Browne, Wolfgang D{\"u}r, Robert Raussendorf, and
  Maarten Van~den Nest.
\newblock Measurement-based quantum computation.
\newblock {\em Nature Physics}, 5(1):19--26, 2009.

\bibitem{raussendorf2012quantum}
Robert Raussendorf and Tzu-Chieh Wei.
\newblock Quantum computation by local measurement.
\newblock {\em Annu. Rev. Condens. Matter Phys.}, 3(1):239--261, 2012.

\bibitem{shor1999polynomial}
Peter~W Shor.
\newblock Polynomial-time algorithms for prime factorization and discrete
  logarithms on a quantum computer.
\newblock {\em SIAM review}, 41(2):303--332, 1999.

\bibitem{grover1996fast}
Lov~K Grover.
\newblock A fast quantum mechanical algorithm for database search.
\newblock In {\em Proceedings of the twenty-eighth annual ACM symposium on
  Theory of computing}, pages 212--219, 1996.

\bibitem{Simonyan2014VeryRecognition}
Karen Simonyan and Andrew Zisserman.
\newblock Very deep convolutional networks for large-scale image recognition.
\newblock In {\em International Conference on Learning Representations}, 2015.

\bibitem{Szegedy2014GoingConvolutions}
Christian Szegedy, Wei Liu, Yangqing Jia, Pierre Sermanet, Scott Reed, Dragomir
  Anguelov, Dumitru Erhan, Vincent Vanhoucke, and Andrew Rabinovich.
\newblock Going deeper with convolutions.
\newblock In {\em Proceedings of the IEEE conference on computer vision and
  pattern recognition}, pages 1--9, 2015.

\bibitem{Voulodimos2018DeepReview}
Athanasios Voulodimos, Nikolaos Doulamis, Anastasios Doulamis, and Eftychios
  Protopapadakis.
\newblock {Deep Learning for Computer Vision: A Brief Review}.
\newblock {\em Computational Intelligence and Neuroscience}, 2018:1--13, 2018.

\bibitem{Sutskever2014SequenceNetworks}
Ilya Sutskever, Oriol Vinyals, and Quoc~V Le.
\newblock Sequence to sequence learning with neural networks.
\newblock In {\em Advances in neural information processing systems}, pages
  3104--3112, 2014.

\bibitem{vamathevan2019applications}
Jessica Vamathevan, Dominic Clark, Paul Czodrowski, Ian Dunham, Edgardo Ferran,
  George Lee, Bin Li, Anant Madabhushi, Parantu Shah, Michaela Spitzer, et~al.
\newblock Applications of machine learning in drug discovery and development.
\newblock {\em Nature Reviews Drug Discovery}, 18(6):463--477, 2019.

\bibitem{biamonte2017quantum}
Jacob Biamonte, Peter Wittek, Nicola Pancotti, Patrick Rebentrost, Nathan
  Wiebe, and Seth Lloyd.
\newblock Quantum machine learning.
\newblock {\em Nature}, 549(7671):195--202, 2017.

\bibitem{wittek2014quantum}
Peter Wittek.
\newblock {\em Quantum machine learning: what quantum computing means to data
  mining}.
\newblock Academic Press, 2014.

\bibitem{schuld2019machine}
Maria Schuld.
\newblock Machine learning in quantum spaces, 2019.

\bibitem{dunjko2018machine}
Vedran Dunjko and Hans~J Briegel.
\newblock Machine learning \& artificial intelligence in the quantum domain: a
  review of recent progress.
\newblock {\em Reports on Progress in Physics}, 81(7):074001, 2018.

\bibitem{aimeur2013quantum}
Esma A{\"\i}meur, Gilles Brassard, and S{\'e}bastien Gambs.
\newblock Quantum speed-up for unsupervised learning.
\newblock {\em Machine Learning}, 90(2):261--287, 2013.

\bibitem{otterbach2017unsupervised}
JS~Otterbach, R~Manenti, N~Alidoust, A~Bestwick, M~Block, B~Bloom, S~Caldwell,
  N~Didier, E~Schuyler Fried, S~Hong, et~al.
\newblock Unsupervised machine learning on a hybrid quantum computer.
\newblock {\em arXiv preprint arXiv:1712.05771}, 2017.

\bibitem{wiebe2014quantum}
Nathan Wiebe, Ashish Kapoor, and Krysta Svore.
\newblock Quantum algorithms for nearest-neighbor methods for supervised and
  unsupervised learning.
\newblock {\em arXiv preprint arXiv:1401.2142}, 2014.

\bibitem{lloyd2013quantum}
Seth Lloyd, Masoud Mohseni, and Patrick Rebentrost.
\newblock Quantum algorithms for supervised and unsupervised machine learning.
\newblock {\em arXiv preprint arXiv:1307.0411}, 2013.

\bibitem{kerenidis2019q}
Iordanis Kerenidis, Jonas Landman, Alessandro Luongo, and Anupam Prakash.
\newblock q-means: A quantum algorithm for unsupervised machine learning.
\newblock In {\em Advances in Neural Information Processing Systems}, pages
  4134--4144, 2019.

\bibitem{schuld2018supervised}
Maria Schuld and Francesco Petruccione.
\newblock {\em Supervised learning with quantum computers}, volume~17.
\newblock Springer, 2018.

\bibitem{benedetti2019parameterized}
Marcello Benedetti, Erika Lloyd, Stefan Sack, and Mattia Fiorentini.
\newblock Parameterized quantum circuits as machine learning models.
\newblock {\em Quantum Science and Technology}, 4(4):043001, 2019.

\bibitem{sergioli2019new}
Giuseppe Sergioli, Roberto Giuntini, and Hector Freytes.
\newblock A new quantum approach to binary classification.
\newblock {\em PloS one}, 14(5):e0216224, 2019.

\bibitem{rebentrost2014quantum}
Patrick Rebentrost, Masoud Mohseni, and Seth Lloyd.
\newblock Quantum support vector machine for big data classification.
\newblock {\em Physical review letters}, 113(13):130503, 2014.

\bibitem{farhi2018classification}
Edward Farhi and Hartmut Neven.
\newblock Classification with quantum neural networks on near term processors.
\newblock {\em arXiv preprint arXiv:1802.06002}, 2018.

\bibitem{schuld2020circuit}
Maria Schuld, Alex Bocharov, Krysta~M Svore, and Nathan Wiebe.
\newblock Circuit-centric quantum classifiers.
\newblock {\em Physical Review A}, 101(3):032308, 2020.

\bibitem{havlivcek2019supervised}
Vojt{\v{e}}ch Havl{\'\i}{\v{c}}ek, Antonio~D C{\'o}rcoles, Kristan Temme,
  Aram~W Harrow, Abhinav Kandala, Jerry~M Chow, and Jay~M Gambetta.
\newblock Supervised learning with quantum-enhanced feature spaces.
\newblock {\em Nature}, 567(7747):209--212, 2019.

\bibitem{mitarai2018quantum}
Kosuke Mitarai, Makoto Negoro, Masahiro Kitagawa, and Keisuke Fujii.
\newblock Quantum circuit learning.
\newblock {\em Physical Review A}, 98(3):032309, 2018.

\bibitem{lecun2015deep}
Yann LeCun, Yoshua Bengio, and Geoffrey Hinton.
\newblock Deep learning.
\newblock {\em nature}, 521(7553):436--444, 2015.

\bibitem{krizhevsky2012imagenet}
Alex Krizhevsky, Ilya Sutskever, and Geoffrey~E Hinton.
\newblock Imagenet classification with deep convolutional neural networks.
\newblock In {\em Advances in neural information processing systems}, pages
  1097--1105, 2012.

\bibitem{grant2018hierarchical}
Edward Grant, Marcello Benedetti, Shuxiang Cao, Andrew Hallam, Joshua Lockhart,
  Vid Stojevic, Andrew~G Green, and Simone Severini.
\newblock Hierarchical quantum classifiers.
\newblock {\em npj Quantum Information}, 4(1):1--8, 2018.

\bibitem{liu2019hybrid}
Junhua Liu, Kwan~Hui Lim, Kristin~L Wood, Wei Huang, Chu Guo, and He-Liang
  Huang.
\newblock Hybrid quantum-classical convolutional neural networks.
\newblock {\em arXiv preprint arXiv:1911.02998}, 2019.

\bibitem{lu2020quantum}
Sirui Lu, Lu-Ming Duan, and Dong-Ling Deng.
\newblock Quantum adversarial machine learning.
\newblock {\em Physical Review Research}, 2(3):033212, 2020.

\bibitem{du2020learnability}
Yuxuan Du, Min-Hsiu Hsieh, Tongliang Liu, Shan You, and Dacheng Tao.
\newblock On the learnability of quantum neural networks.
\newblock {\em arXiv preprint arXiv:2007.12369}, 2020.

\bibitem{huang2020experimental}
He-Liang Huang, Yuxuan Du, Ming Gong, Youwei Zhao, Yulin Wu, Chaoyue Wang,
  Shaowei Li, Futian Liang, Jin Lin, Yu~Xu, et~al.
\newblock Experimental quantum generative adversarial networks for image
  generation.
\newblock {\em arXiv preprint arXiv:2010.06201}, 2020.

\bibitem{schuld2019quantum}
Maria Schuld and Nathan Killoran.
\newblock Quantum machine learning in feature hilbert spaces.
\newblock {\em Physical review letters}, 122(4):040504, 2019.

\bibitem{chopra2005learning}
Sumit Chopra, Raia Hadsell, and Yann LeCun.
\newblock Learning a similarity metric discriminatively, with application to
  face verification.
\newblock In {\em 2005 IEEE Computer Society Conference on Computer Vision and
  Pattern Recognition (CVPR'05)}, volume~1, pages 539--546. IEEE, 2005.

\bibitem{lloyd2020quantum}
Seth Lloyd, Maria Schuld, Aroosa Ijaz, Josh Izaac, and Nathan Killoran.
\newblock Quantum embeddings for machine learning.
\newblock {\em arXiv preprint arXiv:2001.03622}, 2020.

\bibitem{schuld2014quest}
Maria Schuld, Ilya Sinayskiy, and Francesco Petruccione.
\newblock The quest for a quantum neural network.
\newblock {\em Quantum Information Processing}, 13(11):2567--2586, 2014.

\bibitem{rebentrost2018quantum}
Patrick Rebentrost, Thomas~R Bromley, Christian Weedbrook, and Seth Lloyd.
\newblock Quantum hopfield neural network.
\newblock {\em Physical Review A}, 98(4):042308, 2018.

\bibitem{cong2019quantum}
Iris Cong, Soonwon Choi, and Mikhail~D Lukin.
\newblock Quantum convolutional neural networks.
\newblock {\em Nature Physics}, 15(12):1273--1278, 2019.

\bibitem{beer2020training}
Kerstin Beer, Dmytro Bondarenko, Terry Farrelly, Tobias~J Osborne, Robert
  Salzmann, Daniel Scheiermann, and Ramona Wolf.
\newblock Training deep quantum neural networks.
\newblock {\em Nature communications}, 11(1):1--6, 2020.

\bibitem{giovannetti2008quantum}
Vittorio Giovannetti, Seth Lloyd, and Lorenzo Maccone.
\newblock Quantum random access memory.
\newblock {\em Physical review letters}, 100(16):160501, 2008.

\bibitem{perez2020data}
Adri{\'a}n P{\'e}rez-Salinas, Alba Cervera-Lierta, Elies Gil-Fuster, and
  Jos{\'e}~I Latorre.
\newblock Data re-uploading for a universal quantum classifier.
\newblock {\em Quantum}, 4:226, 2020.

\bibitem{adhikary2020entanglement}
Soumik Adhikary.
\newblock An entanglement enhanced training algorithm for supervised quantum
  classifiers.
\newblock {\em arXiv preprint arXiv:2006.13302}, 2020.

\bibitem{cao2020cost}
Shuxiang Cao, Leonard Wossnig, Brian Vlastakis, Peter Leek, and Edward Grant.
\newblock Cost-function embedding and dataset encoding for machine learning
  with parametrized quantum circuits.
\newblock {\em Physical Review A}, 101(5):052309, 2020.

\bibitem{fisher1936use}
Ronald~A Fisher.
\newblock The use of multiple measurements in taxonomic problems.
\newblock {\em Annals of eugenics}, 7(2):179--188, 1936.

\bibitem{anderson1936species}
Edgar Anderson.
\newblock The species problem in iris.
\newblock {\em Annals of the Missouri Botanical Garden}, 23(3):457--509, 1936.

\bibitem{bergholm2018pennylane}
Ville Bergholm, Josh Izaac, Maria Schuld, Christian Gogolin, Carsten Blank,
  Keri McKiernan, and Nathan Killoran.
\newblock Pennylane: Automatic differentiation of hybrid quantum-classical
  computations.
\newblock {\em arXiv preprint arXiv:1811.04968}, 2018.

\bibitem{Tieleman2012}
T.~Tieleman and G.~Hinton.
\newblock {Lecture 6.5---{RmsProp:} {Divide} the gradient by a running average
  of its recent magnitude}.
\newblock COURSERA: Neural Networks for Machine Learning, 2012.

\bibitem{cincio2018learning}
Lukasz Cincio, Yi{\u{g}}it Suba{\c{s}}{\i}, Andrew~T Sornborger, and Patrick~J
  Coles.
\newblock Learning the quantum algorithm for state overlap.
\newblock {\em New Journal of Physics}, 20(11):113022, 2018.

\bibitem{blank2020quantum}
Carsten Blank, Daniel~K Park, June-Koo~Kevin Rhee, and Francesco Petruccione.
\newblock Quantum classifier with tailored quantum kernel.
\newblock {\em npj Quantum Information}, 6(1):1--7, 2020.

\bibitem{preskill2018quantum}
John Preskill.
\newblock Quantum computing in the nisq era and beyond.
\newblock {\em Quantum}, 2:79, 2018.

\bibitem{temme2017error}
Kristan Temme, Sergey Bravyi, and Jay~M Gambetta.
\newblock Error mitigation for short-depth quantum circuits.
\newblock {\em Physical review letters}, 119(18):180509, 2017.

\bibitem{endo2018practical}
Suguru Endo, Simon~C Benjamin, and Ying Li.
\newblock Practical quantum error mitigation for near-future applications.
\newblock {\em Physical Review X}, 8(3):031027, 2018.

\bibitem{Chen2019DetectorTO}
Yanzhu Chen, Maziar Farahzad, Shinjae Yoo, and Tzu-Chieh Wei.
\newblock Detector tomography on ibm 5-qubit quantum computers and mitigation
  of imperfect measurement.
\newblock {\em arXiv: Quantum Physics}, 2019.

\bibitem{bravyi2020mitigating}
Sergey Bravyi, Sarah Sheldon, Abhinav Kandala, David~C Mckay, and Jay~M
  Gambetta.
\newblock Mitigating measurement errors in multi-qubit experiments.
\newblock {\em arXiv preprint arXiv:2006.14044}, 2020.

\end{thebibliography}
\bibliographystyle{unsrt}

\end{document}